\documentclass{article}

\usepackage{arxiv}

\usepackage[utf8]{inputenc} % allow utf-8 input
\usepackage[T1]{fontenc}    % use 8-bit T1 fonts
\usepackage{hyperref}       % hyperlinks
\usepackage{url}            % simple URL typesetting
\usepackage{booktabs}       % professional-quality tables
\usepackage{amsfonts}       % blackboard math symbols
\usepackage{nicefrac}       % compact symbols for 1/2, etc.
\usepackage{microtype}      % microtypography
\usepackage{graphicx}
\usepackage[authoryear,round,sort]{natbib}
\usepackage{doi}
\RequirePackage{float}
\RequirePackage{mathtools}
\RequirePackage{xltabular}
\RequirePackage{booktabs}
\usepackage{caption}
\RequirePackage{subcaption}
\RequirePackage{threeparttable}
\usepackage{xcolor,colortbl}
\usepackage{tablefootnote}
\usepackage{bm}
\usepackage{optidef}
\usepackage{algorithm2e}
\RestyleAlgo{ruled}
\usepackage{hyperref}
\usepackage{amsthm}

\theoremstyle{definition}
\newtheorem{example}{Example}

\title{An Application of D-vine Regression for the Identification of Risky Flights in Runway Overrun}

\author{ {Hassan H.~Alnasser} \\
	Faculty of Mathematics\\
	Technical University of Munich\\
	\texttt{h.alnasser@tum.de} \\
	\And
	{Claudia Czado} \\
	Faculty of Mathematics\\
	Technical University of Munich\\
	\texttt{cczado@ma.tum.de} \\
}

\renewcommand{\shorttitle}{Identification of Risky Flights to Runway Overrun}

\hypersetup{
pdftitle={An Application of D-vine Regression for the Identification of Risky Flights in Runway Overrun},
pdfsubject={Preprint (Risky Flights)},
pdfauthor={Hassan H.~Alnasser, Claudia Czado},
pdfkeywords={Conditional probability estimation, D-vine regression, non Gaussian dependency, runway overrun, risky flight identification},
}

\begin{document}
\maketitle

\begin{abstract}
	In aviation safety, runway overruns are of great importance because they are the most frequent type of landing accidents. Identification of factors which contribute to the occurrence of runway overruns can help mitigate the risk and prevent such accidents. Methods such as physics-based and statistical-based models were proposed in the past to estimate runway overrun probabilities. However, they are either costly or require experts' knowledge. We propose a statistical approach to quantify the risk probability of an aircraft to exceed a threshold at the speed of 80 knots given a set of influencing factors. This copula based D-vine regression approach is used because it allows for complex tail dependence and is computationally tractable. Data obtained from the Quick Access Recorder (QAR) for 711 flights are analyzed. We identify 41 flights with an estimated risk probability $>10^{-3}$ for a chosen threshold and rank the effects of each influencing factor for these flights. Also, the complex dependency patterns between some influencing factors for the 41 flights are shown to be non symmetric. The D-vine regression approach, compared to physics-based and statistical-based approaches, has an analytical solution, is not simulation based and can be used to estimate very small or large probabilities efficiently. 
\end{abstract}

% keywords can be removed

%% Introduction
\section{Introduction} \label{sec: introduction}

The International Air Transport Association (\emph{IATA}) expects the number of air passengers to recover in 2024, exceeding the pre-COVID-19 level by 3 \%. Long term forecast, also, shows a gradual increase in the number of passengers (domestically and internationally) post 2024 \citep{IATA2024}. Therefore, the need for improved safety procedures is essential. There has been an overall decrease in the number of accidents related to commercial aviation in the last 50 years, however, the consequences of such accidents can still result in losses of lives and huge economic costs. Hence, this motivates aviation companies and international aviation agencies to identify and evaluate various risks leading to such accidents. In 2006, the International Civil Aviation Organization (\emph{ICAO}) published the first risk management guidelines \citep{ICAO2006,KeystoSafetyArrival}. Newer guidelines were published and widely accepted by air transport authorities and aviation manufacturers. However, the number of accidents did not decrease, especially for runway overrun accidents \citep{flight}. This encourages aviation safety oversight authorities to move towards a more proactive approach to identify and predict safety related trends \citep{ICAO2013}.

There is a considerable interest to mitigate the risk of runway excursions classified by \emph{IATA}, such as undershoot, veer offs and runway overruns. Such excursions account for about 22 \% of all civil aviation accidents between 1959-2019 \citep{zhao2022research}. The increased danger during landing is due to multiple factors such as weather conditions \citep{wong2006quantifying} and decisions pilots must take while landing \citep{you2013effects, wang2014analysis}. For instance, \cite{jenkins2012reducing} list unstabilized approach, tail or crosswind, high speed and poor use of the reverse thrust as relevant causes for runway overruns. Other authors \citep{chang2016human, ahmed2014real} focus on runway and weather conditions as risk contributors. For example, a large approach speed deviation and runway surface conditions were identified as contributing factors in the Fokker 100 overrun at Newman, Australia in January 2020 \citep{Fokker100}.

Due to confidentiality reasons, access to flight data from accidents is limited. \cite{valdes2011development} compiled a database of air traffic accidents from official commissions and private entities to assign frequencies to factors contributing to 53 runway overruns. They identified long landing (landing far beyond the runway threshold) to be the riskiest, followed by high access approach speed. In this respect, a long distance to reach the controllable speed of 80 knots during landing is considered a precursor of runway overrun. This distance, we will refer to as distance to controllable speed hereafter, is our focus. In particular, we propose a novel statistical approach to estimate the conditional probability that the distance to controllable speed is exceeding a chosen threshold given a set of risk factor values.

Two approaches in modeling runway excursions have been discussed in the literature: one is based on the construction of a physics-based model allowing for simulation and the other is based on statistical models using appropriate flight data.

In the area of physics-based models, \cite{drees2014using} and \cite{Drees} compiled a list of risk factors related to pilot operations and environmental conditions. These factors are  expected to have an influence on the probability of runway overrun and are used as input values for a deterministic physical model. Based on flight dynamics, the physical model is utilized to quantify the associated runway distance to controllable speed. Also, a statistical distribution for each risk factor is estimated using operational flight data obtained from the Quick Access Recorder (\emph{QAR}). These distributions, then, are used for the simulation of appropriate input values to the physical model. A simulation-based approach is needed since a large distance to controllable speed is often not observed in \emph{QAR} based data. With simulated input values, the physical model can be used to generate the associated runway distance to controllable speed for each input value. Thus, to quantify the associated risk probability, one can count the number of simulations which yields a runway distance to controllable speed over a chosen critical threshold. 

To reduce the number of simulations to obtain a nonzero risk probability estimate, the subset simulation approach of \cite{AU2001263} was used. Further, \cite{Drees} proposed, for validation, the comparison of the observed distance to controllable speed from the \emph{QAR} data to the physical model output. \cite{PCE}, however, noted a shift to the left in the distribution of the distance to controllable speed compared to the observed one. This indicates a bias in the model output resulting from either the physical model or the distribution fitting error of the risk factors. Hence, \cite{PCE} proposed to replace the costly physical model by a faster deterministic surrogate model. The proposed model is based on polynomial chaos expansion (for details see \cite{schobi2019global}) and an optimization approach. The optimization approach is used to calibrate the parameters of the fitted input distributions to the physical model to better match the \emph{QAR} observed distribution. 

Several statistical approaches have also been utilized. For example, unconditional frequency and hierarchical Bayesian models have been considered in \cite{ARNALDOVALDES2018216}. \cite{run2014estimation} used a linear regression model for the landing distance, while \cite{wagner2014statistical} applied logistic regression and its Bayesian version. Both models were used to model the probability of fatalities using 1400 records of runway excursions happened between 1970 and 2009. The problem of hard landings, also, was considered in \cite{hu2016study} where a support vector machine model was applied. Further, discrete Bayesian networks have been utilized. \cite{zwirglmaier2016discretization}, for example, used first order Taylor expansions to perform the necessary discretization of a designed Bayesian network. \cite{ayra2019bayesian} used the graphical network interface (GeNIe) software \citep{GeNIe} which starts with a network proposed by experts. Most recently, \cite{zhao2022research} developed a neural network to model the landing distance.

While the physics-based modeling approach of \cite{Drees} with the calibration of the input parameters suggested by \cite{PCE} allows for the identification of risk conditions, both models do not allow for quantifying the effect of each risk factor having on the occurrence of runway overrun. The statistical models, while not simulation based, have other shortcomings. Those suggested by \cite{wang2014quantification} and \cite{ARNALDOVALDES2018216} are unconditional and, thus, cannot model the influence of several risk factors jointly. This task is achieved with the Bayesian network approaches; however, they require discretization of the continuously measured risk factors. Moreover, Bayesian network approaches are often based on networks using the subjective knowledge by experts. This suggests that there is a room to develop a flexible statistical framework to allow identification of risky flights and quantify the effect of the influencing factors. Such a framework should also be able to model dependency patterns among the variable of interest, given here by the distance to controllable speed, and the set of contributing factors, especially in the tails. 

Therefore, we propose to follow a copula based approach since copulas allow to describe dependence behavior separately from marginal behavior. A copula is a multivariate distribution function with uniform marginal distributions. 

To increase the range of dependence patterns in high dimensions, \cite{joe1997multivariate} and later \cite{Cook&bedf} proposed using conditioning to construct multivariate copulas. This requires only the specification of bivariate copula terms which yields the vine copula class. The seminal paper of \cite{PPC} made this pair copula construction approach applicable. The reader can consult the books by \cite{joe2014dependence} and \cite{czado2019analyzing} as well as the survey by \cite{CzadoNaglerReview} for details and recent developments. The vine copula class is especially suited to model asymmetric tail dependence. 

 We use the D-vine (quantile) regression approach of \cite{KRAUS20171} to achieve our objective. The restriction to the subclass of D-vines allows us to express the conditional distribution function of the variable of interest given the potentially influencing factors analytically. In addition, the D-vine regression is well suited to model extreme large or small conditional probabilities, allowing for tail dependence. In this paper, we show how the D-vine regression based approach can be utilized to identify risky flights, flights having a distance to controllable speed greater than 2500 meters with estimated probability $>10^{-3}$. Among 711 flights, we identify 41 to be risky. All flights were of the same aircraft type and landed at the same airport. Furthermore, we rank the marginal effect of each contributing factor for the risky flights. The ranking, in a descending order, is given as follows: brake duration, headwind speed, time brake started, touchdown, equivalent acceleration and approach speed deviation. In addition, we study the joint behavior for all pairs of contributing factors for the risky flights. This shows dependence among the contributing factors. We show a non-symmetric dependence between time brake started and brake duration as well as between headwind speed and equivalent acceleration. 

We, further, investigate if a standard linear quantile regression approach of \cite{Koenker} and \cite{koenker2001quantile} is able to estimate such small risk probabilities. This is not the case and, even more, quantile crossing occurs in this data set. 

The remainder of this paper is organized as follows. In Section \ref{sec: data descri.}, we provide more details on the \emph{QAR} data used. We introduce and discuss in Section \ref{sec: methodology} the shortcomings of the linear quantile regression approach with regard to estimating conditional probabilities in the tail. Following this, we review D-vine copulas briefly and discuss the D-vine (quantile) regression approach. Estimation of critical event probabilities is then discussed and presented in Section \ref{sec: methodology}. Section \ref{sec: data analysis} presents results from the linear quantile and the D-vine regression. Finally, a brief discussion is given in Section \ref{sec: conclusion}.

%% Data 
\section{Data description} \label{sec: data descri.}
\begin{figure}[H]
		\centering
		\includegraphics[width=1\textwidth]{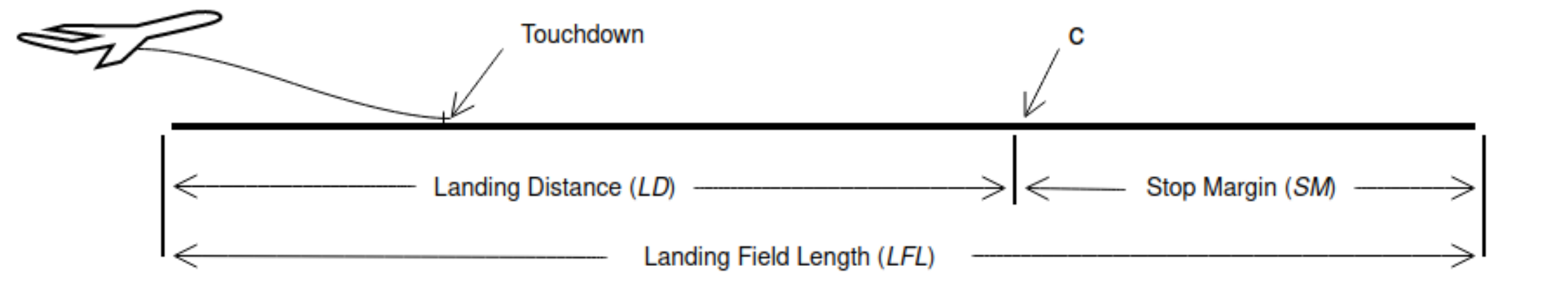}
		\caption{Landing phase illustration.}
		\label{fig: landing illustration}
	\end{figure}

We consider the same data set used in \cite{Drees} and \cite{PCE}. The data consists of 11 continuous variables, referred to as contributing factors, and the distance to controllable speed as the response variable. The response variable is referred to \emph{th80} hereinafter. We list in Table \ref{tab: Contr. Fact.} the contributing factors which correspond to observed and measured parameters potentially leading to runway overrun incidents. It is worth mentioning that \cite{Drees} considers a runway overrun incident when a stop margin (\emph{SM}) is less than zero. A stop margin \nocite{scholz2012aircraft} is the difference between a landing field length (\emph{LFL}) and a landing distance (\emph{LD}), $SM \coloneqq LFL - LD $. See Figure \ref{fig: landing illustration} for illustration.

In addition, we consider the assumption that an aircraft should reach 80 knots (\emph{kts}) ground speed before a fixed threshold $c$, leaving a reasonable \emph{SM}. See Figure \ref{fig: landing illustration} for demonstration. The so-called 80 \emph{kts} ground speed is a controllable and safe speed \citep{Drees}. More specifically, a pilot is able to control the aircraft at the speed of 80 \emph{kts} or less. Thus, this leads to our definition of runway overrun. A runway overrun incident occurs when an aircraft exceeds the fixed threshold $c$ at a speed greater than $80$ \emph{kts} without a reasonable stop margin remained. 

As mentioned above, similar contributing factors have been used to assess the risk of runway overrun in \cite{PCE} and \cite{Drees}. However, they considered air density (\emph{ad}) in place of temperature (\emph{temp}) and reference air pressure (\emph{refAP}).  Air density was defined as $ad = refAP/(R \times temp)$ for a specific gas constant \emph{R}. We use, in our application, the observed values of \emph{temp} and \emph{refAP} instead of the derived \emph{ad}.  
\begin{xltabular}{\linewidth}{ l  X }
  \caption{Contributing factors' definitions and measurement units.} 
 \label{tab: Contr. Fact.}\\
\toprule
 \textbf{Contributing Factor} & \textbf{Definition}  \\
\midrule
\endfirsthead
\toprule
 \textbf{Contributing Factor} & \textbf{Definition}  \\
\midrule
\endhead
\bottomrule
\endfoot

Headwind speed (\textit{hws}) & Headwind speed measured at touchdown (\textit{td}) in $m/s$.\\   \addlinespace

Temperature (\textit{temp}) & Temperature in \emph{Kelvin} provided by \emph{METAR}\footnote{METeorological Aerodrome Report}.\\   \addlinespace

Reference air pressure (\textit{refAP}) & Reference air pressure in \emph{hPa}.\\   \addlinespace

Approach speed deviation (\textit{asd}) & Deviation in speed between target approach speed and the actual true airspeed at \textit{td} in $m/s$. \\  \addlinespace

Time of deploying reversers (\textit{trd}) & Time reversers deployed after \textit{td} in seconds $s$. \\ \addlinespace 

Time of deploying spoilers (\textit{tsd}) & Time spoilers deployed after \textit{td} in $s$.\\ \addlinespace 

Landing mass (\textit{lm}) & Landing weight taken at \textit{td} in $kg$.\\   \addlinespace

Time of starting brake (\textit{tbs}) & Time brakes started after \textit{td} in $s$.\\ \addlinespace  

Duration of braking (\textit{bd}) & Brake duration until $80$ \emph{kts} in $s$.\\  \addlinespace

Touchdown Distance (\textit{td}) & Distance from \textit{th} to \textit{td} in $m$.\\  \addlinespace

Equivalent acceleration (\textit{ea}) & Constant deceleration from \textit{td} to 80 $kts$ in $m/s^{2}$. \\  \addlinespace
\end{xltabular}

\begin{figure}[ht]
		\centering
		\includegraphics[width=0.46\textwidth]{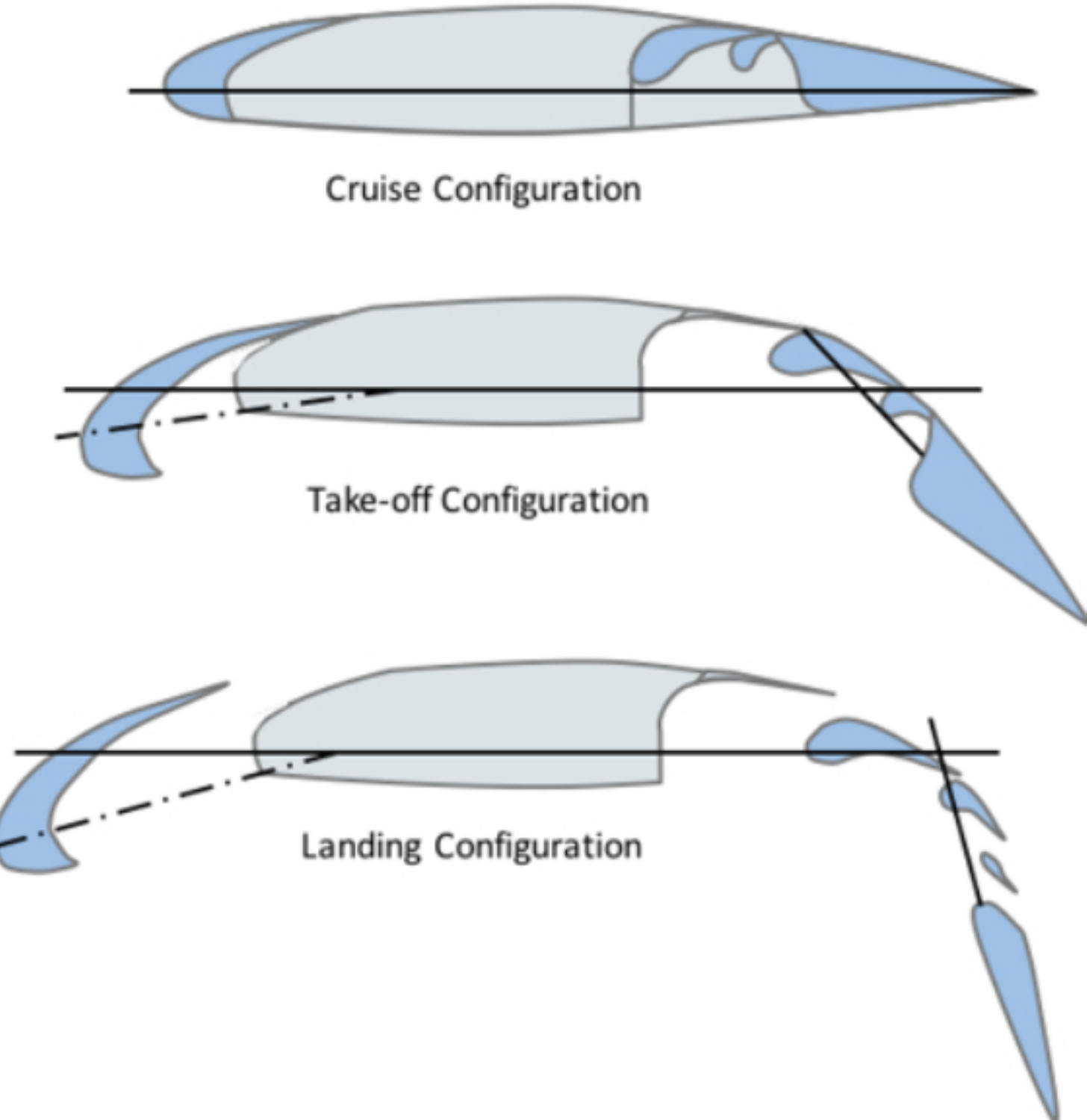}
		\caption{Three different flight phases (cruise, take-off and landing) with different flap and slat configurations \citep{simons2015model}.}
		\label{fig: flight config.}
	\end{figure}

 To insure comparability, we focus on 711 flights of the same aircraft type. All 711 flights landed on the same runway in both directions. Moreover, there are additional discrete factors which we do not include because they have the same setting (or value) for the 711 flights. These factors describe aircraft systems which are implemented when an aircraft is either landing or taking off. Such factors have several configurations. See Table \ref{Tab: aircraft sys} for more details. For example, the combination of fully extended flaps and slats generates more drag (see Figure \ref{fig: flight config.}) and is used to fly slower at a higher power setting\nocite{anderson2007fundamentals} \citep{sforza2014commercial}. We shade in Table \ref{Tab: aircraft sys} the configurations and weather condition we observed in our flight data.  
 
 	\begin{table}[h!]
 	\scalebox{1.14}{
 	\begin{threeparttable}
    \caption{Aircraft systems and their configurations as well as runway condition.}
        \label{Tab: aircraft sys}
            \begin{tabular}{@{}llllll}
                \hline
                    \textbf{flapConfig}\tablefootnote{Flap configuration position at different degrees.}  : & CONF ${0}^{\circ}$ & CONF ${10}^{\circ}$ & CONF ${20}^{\circ}$ & CONF ${25}^{\circ}$ & \cellcolor[HTML]{C0C0C0}CONF ${30}^{\circ}$  \\[0.75ex] 
                    \textbf{slatConfig}\tablefootnote{Slat configuration position at different degrees.}  : & CONF ${0}^{\circ}$ & CONF ${5}^{\circ}$ & CONF ${10}^{\circ}$ & \cellcolor[HTML]{C0C0C0}CONF ${20}^{\circ}$ &  \\ 
                    \textbf{revThrust}\tablefootnote{Reverse thrust either applied fully or partially.}   : & 3/ALL OUT & \cellcolor[HTML]{C0C0C0}FullRev & 2 OUT &  &  \\ 
                    \textbf{splrSysStat}\tablefootnote{Spoiler system status either operative or partially/fully inoperative.} : & \cellcolor[HTML]{C0C0C0}OP  & $\leq$ 2 FAULT & $\leq$ 4 FAULT & 5/ALL FAULT &  \\ 
                    \textbf{brkSysStat}\tablefootnote{Brake system status (operative, degraded, inoperative).} : & \cellcolor[HTML]{C0C0C0}OP  & DEGRADED  & INOP & &  \\ 
                    \textbf{rwyCond}\tablefootnote{Runway condition.}    : & \cellcolor[HTML]{C0C0C0}DRY & WET & VICINITY &  &  \\ 
                    \hline
            \end{tabular}
            \end{threeparttable}}
    \end{table}
	
\section{Methodology} \label{sec: methodology}

We propose a statistical model to estimate risk probabilities for our flight data. The estimated risk probabilities correspond to a flight exceeding a chosen threshold $c$ at the controllable speed of 80 \emph{kts}. However, we first introduce some notations and terminologies, which will help us in defining our model. We, then, review linear quantile regression (\emph{lqr}), our a benchmark model, and discuss some of its disadvantages.     

\subsection*{Data terminology}

Suppose we have observed data $\bm{y}=(y_{1},\ldots,y_{n})^\top$ composed of $n$ independent observations. Each observation $y_i$ (${i = 1,\ldots,n}$) has $d$ covariates collected in the covariate vector $\bm{x}_i$ such that $\bm{x}_i = (x_{i1}, \ldots, x_{id})^\top$.

\subsection{Linear quantile regression} \label{subsec: lqr}
When it comes to methods to predict conditional quantiles, the literature is rich with such methods. The first and probably the most widely used is the \emph{lqr} \citep{Koenker, koenker2001quantile}. Further methods followed after such as local quantile regression \citep{spokoiny2013local}, semiparametric quantile regression \citep{noh2015semiparametric} and nonparametric qunatile regression \citep{li2013optimal}. While most of the recent extensions to \emph{lqr} are nonparametric \citep{koenker2017handbook}, we use the parametric \emph{lqr} of \cite{koenker_2005}. This is due to the linear relationship between \emph{th80} and the contributing factors. See Figure \ref{fig: pairwise-raw} in Appendix \ref{AppB} for illustration.

\cite{Koenker} introduced the parametric \emph{lqr} as a method to estimate conditional quantiles of a response variable $Y$ given \emph{d} covariates, $\bm{x} = (x_1, \ldots, x_d)^\top.$ More specifically, the conditional quantile function is defined as: 
    \begin{equation}
        q_{\alpha}^{(l)}(x_1,\ldots,x_d):=F^{-1}_{Y|\bm{X}}(\alpha|x_1,\ldots,x_d),
        \label{eq: lqr cond. quan.}
    \end{equation}
    where $F_{Y|\bm{X}}$ is the conditional distribution function of $Y$ given $\bm{X} = \bm{x}$ and $\alpha \in (0,1)$ is the quantile level. It is important to note that the \emph{lqr} estimator is more robust in the presence of non-normal errors and outliers \citep{hao2007quantile}. In addition, the \emph{lqr} gives a more complete representation of the conditional response distribution for a set of $\alpha$ levels. 
    
    Originally, \cite{Koenker} assumed a linear model for the conditional quantile function given by
    \begin{equation}
	    q_{\alpha}^{(l)}\left(x_{i1},\ldots,x_{id}\right)=\beta_0+\sum_{j=1}^d{\beta_j x_{ij}},
	    \label{eq: lqr cond. quant. reg}
    \end{equation}
where $\bm{\beta}$ $\in$ $\mathbb{R}^{d+1}$ represents the unknown regression coefficients. 
These coefficients are estimated by solving the minimization problem:
    \begin{equation}
        \begin{aligned}
            \min_{\bm{\beta}\in\mathbb{R}^{d+1}}\left\{ \alpha\sum_{i=1}^n{\left(y_{i}-\beta_0-\sum_{j=1}^d{\beta_j x_{ij}}\right)}+(1-\alpha)\sum_{i=1}^n{\left(\beta_0+\sum_{j=1}^d{\beta_j x_{ij}}-y_{i}\right)}\right\}.  
        \end{aligned}\\
        \label{eq: lqr minimization}
    \end{equation}
    
\begin{figure}[H]
		\centering
		\includegraphics[width=.65\textwidth]{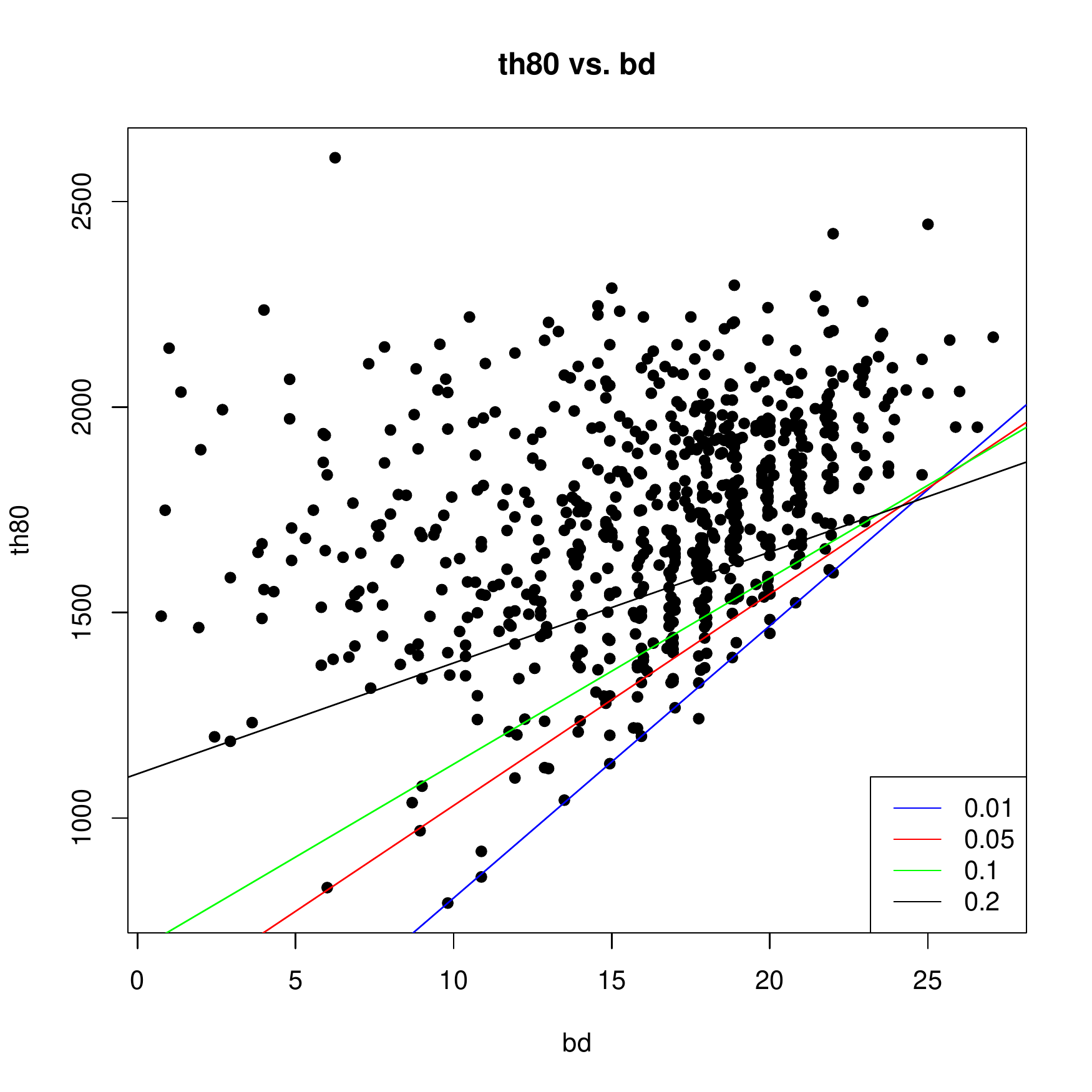}
		\caption{Fitted quantile regression lines for \emph{th80} vs. \emph{bd} at $\alpha = (0.01, 0.05, 0.1, 0.2)^\top$.}
		\label{fig: lqr crossings}
	\end{figure}

Although \emph{lqr} is a useful approach without any distributional assumptions, it has a major flaw in that the regression lines of several quantile levels may cross. Figure \ref{fig: lqr crossings} shows four fitted quantile regression lines crossing using the observed \emph{QAR} data. The cause of this is the piecewise linear formulation of the check function in (\ref{eq: lqr minimization}). \cite{BernCzado}, additionally, show that Equation (\ref{eq: lqr cond. quant. reg}) is only satisfied when the dependent variable and covariates ($Y,$ $\bm{X}$) are jointly multivariate normally distributed.

In the next subsections, we introduce the concept of D-vine copulas and D-vine copula based quantile regression. The D-vine regression approach overcomes the mentioned pitfalls of the \emph{lqr}. 

%In Section \ref{subsec: d-vine regression} we introduce a statistical copula based model that overcomes the %pitfalls of \emph{lqr}. For this, we first briefly discuss the concept of a \emph{copula} and introduce %$D$-vine copula. This will form the backbone of our risk assessment approach. 

\subsection{\textit{D}-vine copulas} \label{subsec: d-vine copula}

Modeling the dependence structure of multivariate data is an important step in predictive modeling. Copulas have been used successfully in modeling complex dependence structures in multiple disciplines, such as geotechnical engineering \citep{tang2015copula} and financial time series \citep{patton2012review}. 

A \textit{d}-dimensional copula \emph{C} is a \textit{d}-variate distribution function defined on the unit hypercube $[0,1]^{\textit{d}}$ with uniformly distributed margins. \cite{Skla59} showed that for every multivariate random vector $\bm{X}=(X_1,\ldots,X_d)^\top$ with joint distribution function  $F$ and marginal distribution functions $F_1, \ldots, F_d$, there exists a copula $C$, such that
    \begin{equation}
        F(x_{1},\ldots,x_{d})=C(F_{1}(x_1),\ldots,F_{d}(x_d)),
        \label{eq: sklar's}
    \end{equation}
for $\bm{x} \in \mathbb{R}^{d}$. The copula $C$ is unique when $\bm{X}$ is absolutely continuous. An advantage of the copula approach is that it allows for separation of the marginal behavior of each variable from the joint dependence structure specified by the copula \emph{C}. This is different from the standard approach, \emph{i.e.} multivariate normal, where all margins and the joint distribution are assumed normal and, thus, only allow for symmetric dependence. 

To obtain uniformly distributed margins, $U_j \coloneqq F_j(X_j)$, we apply the Probability Integral Transform (\emph{PIT}), $F_{j}(\cdot),$ to each variable $X_j$, $j = 1, \ldots, d.$ Consequently, this allows us to write the joint density function $f$ of Equation (\ref{eq: sklar's}) as a product of the copula density $c$ and the marginal densities as
    \begin{equation}
        f(x_{1},\ldots,x_{d})=c(F_{1}(X_1),\ldots,F_{d}(X_d)) \cdot f_1(x_1) \ldots f_d(x_d).
        \label{eq: copula joint density}
    \end{equation}

Note that the copula approach permits us to specify arbitrary margins. $X_j$ can, for example, be Gamma or Beta distributed. The choice of multivariate copula families was limited with regard to the allowable degree of asymmetric tail dependence in the past. To increase the modeling flexibility, \cite{joe1997multivariate} constructed multivariate copulas from bivariate copulas. The idea of conditioning was utilized to build a valid multivariate copula distribution using the bivariate copulas. Since there are multiple ways to select the needed conditioning variables, \cite{Cook&bedf} proposed a graphical structure called a vine tree structure to identify the different choices for the conditioning variables. This led to the birth of the regular (R)-vine copula class, which is the topic of recent books (\cite{joe2014dependence, czado2019analyzing}).  

The regular vine tree structure consists of a set of linked trees ($T_1, \ldots, T_{d-1}$) where the nodes in the first tree $T_1$ are the variables $X_1, \ldots, X_d$. The edges from $T_1$ will become the nodes of $T_2$, and any edge of tree $T_2$ is allowed as long as the nodes of $T_2$ share a node in $T_1$. This tree construction principle is carried forward for the remaining trees and is called the proximity condition. Now, each edge of a tree is associated with a bivariate pair copula. The product of all these pair copula densities, which are evaluated at conditional distribution functions, is then a valid joint copula density. The modeling potential, including step-wise estimation approaches, was first discussed in \cite{PPC}. This is also where the term Pair Copula Construction (\emph{PCC}) for constructing vine copulas was used. 

An earlier review on the use of \emph{PCC} in financial applications was given in  \cite{aas2016pair}, and in Chapter 11 of \cite{czado2019analyzing} further applications in the engineering and life sciences domains were presented. More recently, \citet{CzadoNaglerReview} gives an overview of vine copula based modeling. 

Two popular sub-classes of regular vines are the canonical (C)-vines and drawable (D)-vines. For our conditional risk assessment approach, we will use D-vines. The D-vine quantile regression, first proposed by \citet{KRAUS20171}, models the dependence between the response and covariates flexibly and also allows for a forward variable selection. These attributes make the D-vine approach suitable for our application since we want to model the distance to controllable speed conditioned on a set of risk factors flexibly.    

The following notations are needed to describe the conditional distributions in the D-vine class. For a $d$-dimensional random vector $\bm{X}$, let a set $\mathcal{D} \subset \left\{1,\ldots,d\right\}$ such that $\bm{X}_{\mathcal{D}}$ is a sub random vector and $\bm{x}_{\mathcal{D}}$ is its value. For $i,j\in\left\{1,\ldots,d\right\} \backslash \mathcal{D},$ we define:
\begin{itemize}
	\item $C_{X_i,X_j;\bm{X}_\mathcal{D}}(\cdot,\cdot;\bm{x}_\mathcal{D})$ is the bivariate copula associated with the conditional distribution of $(X_i,X_j)$ given $\bm{X}_\mathcal{D}=\bm{x}_\mathcal{D}$. We use the following abbreviation  $C_{ij;\mathcal{D}}(\cdot,\cdot;\bm{x}_\mathcal{D})$ and $c_{ij;\mathcal{D}}(\cdot,\cdot;\bm{x}_\mathcal{D})$ for the distribution function and density, respectively.
	\item $F_{X_i|\bm{X}_\mathcal{D}}(\cdot|\bm{x}_\mathcal{D})$ is the univariate conditional distribution of the random variable $X_i$ given $\bm{X}_\mathcal{D}=\bm{x}_\mathcal{D}$, which is abbreviated by $F_{i|\mathcal{D}}(\cdot|\bm{x}_\mathcal{D})$.
	\item $C_{U_i|\boldsymbol{U}_\mathcal{D}}(\cdot|\boldsymbol{u}_\mathcal{D})$ is the conditional distribution of the \emph{PIT} random variable $U_i$ given $\boldsymbol{U}_\mathcal{D}=\boldsymbol{u}_\mathcal{D}$, which is abbreviated by $C_{i|\mathcal{D}}(\cdot|\boldsymbol{u}_\mathcal{D})$.
\end{itemize}

\cite{czado2010pair} expresses the joint density $f$ in the case of a D-vine distribution as:
\begin{multline}
	f(x_1,\ldots,x_d) = \prod_{k=1}^d{f_k(x_k)}\prod_{i=1}^{d-1}\prod_{j=i+1}^dc_{ij;i+1,\ldots,j-1}\big(F_{i|i+1,\ldots,j-1}\left(x_i|x_{i+1},\ldots,x_{j-1}\right),\\
	F_{j|i+1,\ldots,j-1}\left(x_j|x_{i+1},\ldots,x_{j-1}\right);x_{i+1},\ldots,x_{j-1}\big),\\
	\label{eq: d-vine-density-indices}
\end{multline}
for distinct indices $ i $ and $j$, $i < j.$

\begin{figure}[H]
		\centering
		\includegraphics[width=0.55\textwidth]{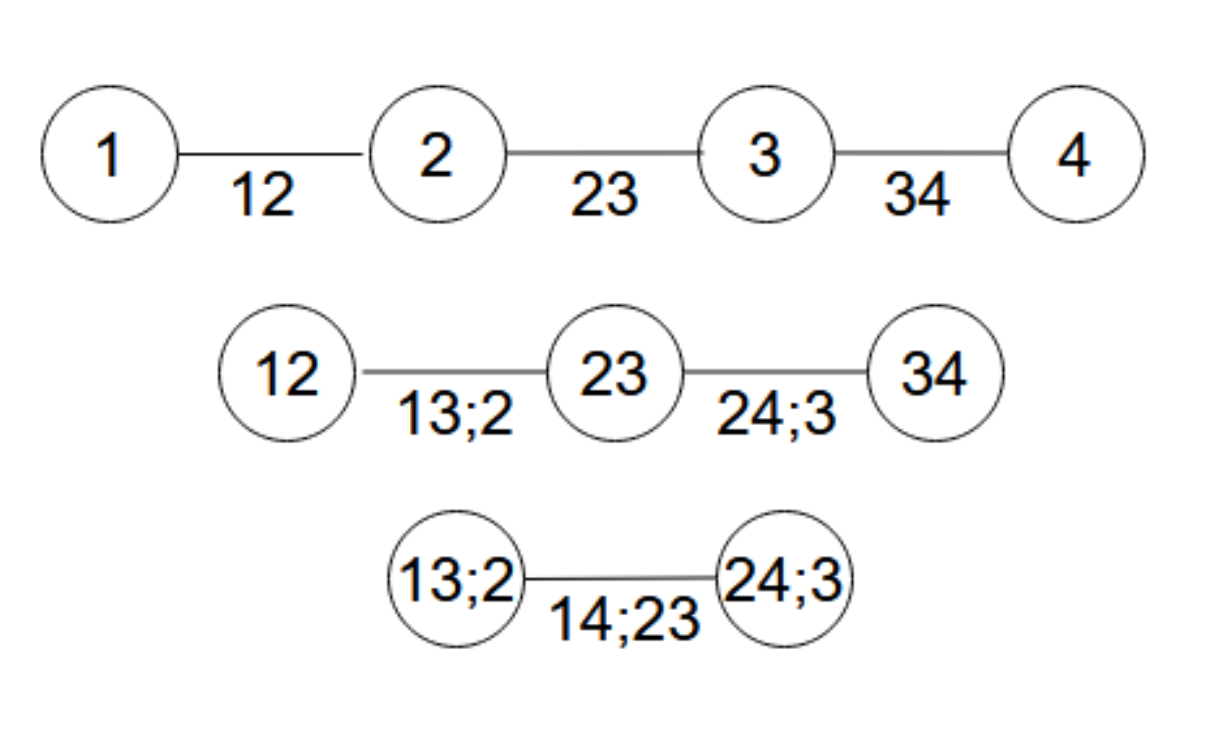}
		\caption{Graphical representation of a four-dimensional D-vine with order $X_1-X_2-X_3-X_4$.}
		\label{fig: 4-d-vine-ex}
	\end{figure}

An illustration of a four-dimensional D-vine distribution is given in Example \ref{ex: 4-d-vine}, and its graphical representation is shown in Figure \ref{fig: 4-d-vine-ex}. 
\begin{example}[Four-dimensional D-vine] \label{ex: 4-d-vine}
A D-vine distribution for $d = 4$ has a joint density given by 
    \begin{align*}
        f(x_{1}, x_{2}, x_{3}, x_{4}) = & f_4(x_4) f_3(x_3)  f_2(x_2)  f_1(x_1) &&\\
        &   c_{12}(F_1(x_1), F_2(x_2)) \cdot c_{23}(F_2(x_2), F_3(x_3)) \cdot c_{34}(F_3(x_3), F_4(x_4)) && (T_{1}) \\
        &   c_{13;2}(F_{1|2}(x_{1}|x_{2}), F_{3|2}(x_3|x_2)) \cdot c_{24;3}(F_{2|3}(x_{2}|x_{3}), F_{4|3}(x_4|x_3)) &&  (T_{2}) \\
        &   c_{14;23}(F_{1|23}(x_{1}|x_{2}, x_{3}), F_{4|23}(x_{4}|x_{2},x_{3})). && (T_{3})
    \end{align*}
\end{example}

$f(x_1,x_2,x_3,x_4)$ is decomposed into 
\begin{equation*}
    f_{4|123}(x_4|x_1,x_2,x_3) \cdot f_{3|12}(x_3|x_1,x_2) \cdot f_{2|1}(x_2|x_1) \cdot f_{1}(x_1),
\end{equation*}
where each conditional density is considered separately. We write $f_{3|12}(x_3|x_1,x_2)$ in terms of bivariate copulas and a marginal density using Sklar's theorem (\ref{eq: sklar's}) as follows:
\begin{align} \label{eq: 3-decompos}
    f_{3|12}(x_3|x_1,x_2) &=  \frac{f_{13|2}(x_1,x_3|x_2)}{f_{1|2}(x_1|x_2)} \nonumber \\
    &=  \frac{c_{13;2}(F_{1|2}(x_1|x_2), F_{3|2}(x_3|x_2)) \cdot f_{1|2}(x_1|x_2) \cdot f_{3|2}(x_3|x_2)}{f_{1|2}(x_1|x_2)} \nonumber \\
    &=  c_{13;2}(F_{1|2}(x_{1}|x_{2}), F_{3|2}(x_3|x_2)) \cdot f_{3|2}(x_3|x_2), 
\end{align}
where, further, we decompose $f_{3|2}(x_3|x_2)$ into
\begin{align*}
    f_{3|2}(x_3|x_2) & = \frac{f_{23}(x_2,x_3)}{f_2(x_2)}\\
    & = \frac{c_{23}(F_2(x_2), F_3(x_3)) \cdot f_2(x_2) \cdot f_3(x_3)}{f_2(x_2)} \\
    & = c_{23}(F_2(x_2), F_3(x_3)) \cdot f_3(x_3).
\end{align*}
Similarly, we write $f_{4|123}(x_4|x_1,x_2,x_3)$ as 
 \begin{align*}
        f_{4|123}(x_4|x_1,x_2,x_3) & =  c_{14;23}(F_{1|23}(x_{1}|x_{2}, x_{3}), F_{4|23}(x_{4}|x_{2},x_{3})) \\
        & \cdot  c_{24;3}(F_{2|3}(x_{2}|x_{3}), F_{4|3}(x_4|x_3)) \cdot c_{34}(F_3(x_3), F_4(x_4)) \cdot f_4(x_4).
    \end{align*}

 This gives us the joint density given above in Example \ref{ex: 4-d-vine}, and the corresponding D-vine tree structure is given in Figure \ref{fig: 4-d-vine-ex}. We see the variables are the nodes of tree $T_{1}$, and the edges of tree $T_{1}$ correspond to the non conditional bivariate copula densities in Example \ref{ex: 4-d-vine}. The arrangement of the variables in tree $T_{1}$ is arbitrary, so we denote this specific order by $X_1-X_2-X_3-X_4$. The edges $12$, $23$ and $34$ now become nodes in tree $T_2$. The nodes $12$ and $23$ can be connected by an edge denoted by $13;2$ since the edges $12$ and $23$ share the common node $2$ in tree $T_1$. 

To fit a D-vine copula with a specific order to given data, the pair copulas in Equation (\ref{eq: d-vine-density-indices}) will be estimated parametrically for our application. This allows for easy model simulation. In addition, we assume the copulas associated with the conditional distributions $C_{i,j;\mathcal{D}}$ do not depend on the specific value $\bm{x}_{\mathcal{D}}$ of the conditioning vector $\boldsymbol{X}_{\mathcal{D}}$. This assumption is called the simplifying assumption, and it is assumed in (\ref{eq: d-vine-density-indices}) for tractability reasons.

To evaluate the conditional distributions $F_{i|i+1,\ldots,j-1}\left(x_i|x_{i+1},\ldots,x_{j-1}\right)$ in Equation (\ref{eq: d-vine-density-indices}), we only need pair copulas specified in the lower trees of the D-vine. Together with the simplifying assumption, this allows us to determine them recursively \citep{joe1997multivariate}. In detail, let $ \mathcal{D} = \{i+1, \ldots, j-1\},$ then we can write $F_{i|i+1,\ldots,j-1}\left(x_i|x_{i+1},\ldots,x_{j-1}\right)$ as $F_{i|\mathcal{D}}(x_i|\boldsymbol{x}_{\mathcal{D}}).$ Subsequently, we can express $F_{i|\mathcal{D}}(x_i|\boldsymbol{x}_{\mathcal{D}})$ for $ l \in \mathcal{D}$ and $\mathcal{D}_{-l}:=\mathcal{D}\backslash \left\{l\right\}$ as
\begin{equation}
	F_{i|\mathcal{D}}\left(x_i|\boldsymbol{x}_\mathcal{D}\right) = h_{i|l;\mathcal{D}_{-l}}\left(F_{i|\mathcal{D}_{-l}}\left(x_i|\boldsymbol{x}_{\mathcal{D}_{-l}}\right)|F_{l|\mathcal{D}_{-l}}\left(x_{l}|\boldsymbol{x}_{\mathcal{D}_{-l}}\right)\right),
	\label{eq: joe97}
\end{equation}
where for $i,j\notin \mathcal{D}, i<j$, $h_{i|j;\mathcal{D}}(u|v) \coloneqq {\partial C_{ij;\mathcal{D}}(u,v)}/{\partial v}=C_{i|j;\mathcal{D}}(u|v)$ and, similarly,   $h_{j|i;\mathcal{D}}(v|u)={\partial C_{ij;\mathcal{D}}(u,v)}/{\partial u}=C_{j|i;\mathcal{D}}(v|u)$. These are called $h$-functions, which are associated with the pair-copula $C_{ij;\mathcal{D}}$. Note that the $h$-functions are independent of the specific value $\bm{x}_{\mathcal{D}}$ for $\boldsymbol{X}_{\mathcal{D}}$ because of the simplifying assumption. A general property of vine densities is that all required conditional distribution functions can be determined using only $h$-functions.

%involving pair copulas occurring in the density expression, such as (\ref{eq: d-vine-density-indices}) for D-vines.
\begin{example}[Conditional distribution functions]
We illustrate how to determine conditional distribution functions for a three-dimensional vector $\bm{X}=(X_1,X_2,X_3)^\top$:
\begin{align*}
    F_{3|12}(x_3|x_1,x_2) & = \int_{-\infty}^{x_3} f_{3|12}(t_3|x_1,x_2) \,dt_{3} \\
     & = \int_{-\infty}^{x_3} c_{13;2}(F_{1|2}(x_{1}|x_{2}), F_{3|2}(t_3|x_2)) \cdot f_{3|2}(t_3|x_2) \,dt_{3}\\
     & = \int_{-\infty}^{x_3} \frac{\partial}{\partial F_{1|2}(x_1|x_2)} \frac{\partial}{\partial F_{3|2}(t_3|x_2)} C_{13;2}(F_{1|2}(x_{1}|x_{2}), F_{3|2}(t_3|x_2)) \\
     & \cdot f_{3|2}(t_3|x_2) \,dt_{3}\\
     & = \frac{\partial}{\partial F_{1|2}(x_1|x_2)} \int_{-\infty}^{x_3}\left[\frac{\partial}{\partial t_3} C_{13;2}(F_{1|2}(x_{1}|x_{2}), F_{3|2}(t_3|x_2))\right] \,dt_{3} \\
     & = \frac{\partial}{\partial F_{1|2}(x_1|x_2)} C_{13;2}(F_{1|2}(x_{1}|x_{2}), F_{3|2}(x_3|x_2)) \\
     & = h_{1|3;2}\left( h_{1|2}(x_1 | x_2)| h_{3|2}(x_3|x_2)\right).
\end{align*}
\end{example}

\subsection{\textit{D}-vine copula based quantile regression} \label{subsec: d-vine regression}

The D-vine copula based quantile regression, proposed by \cite{KRAUS20171}, predicts the quantile of a response variable $Y$ given some covariates $X_1, \ldots, X_d$, for $d \geq 1$. For convenience, we use the following notation: $Y \sim F_Y,$ and $X_j \sim F_j,$ $ j = 1, \ldots, d.$ This helps us in defining the conditional quantile function $Y|\boldsymbol{X} = \boldsymbol{x}$ for level $\alpha \in (0,1)$, which is expressed as the inverse of the conditional distribution function of $Y|\boldsymbol{X}$: 
\begin{equation}
    q_{\alpha}(x_1,\ldots,x_d)\coloneqq F^{-1}_{Y|X_1,\ldots,X_d}(\alpha|x_1,\ldots,x_d).
    \label{eq: d-vine-inv-con-dis}
\end{equation}

Applying the \emph{PIT} to the response $Y$ and to the covariates $X_1, \ldots, X_d$, we obtain $ V \coloneqq F_Y(Y)$ and $U_j \coloneqq F_j(X_j)$, respectively. With the corresponding observed \emph{PIT} values $v \coloneqq F_Y(y)$ and $u_j \coloneqq F_j(x_j),$ we have the following: 
    \begin{align}
        F_{Y|X_{1},\ldots, X_{d}}(y|x_{1},\ldots,x_{d})&=P(Y\leq y|X_1=x_{1},\ldots,X_d=x_{d}) \nonumber\\
        &=P(F_Y(Y)\leq v|F_1(X_1)=u_{1},\ldots,F_d(X_d)=u_{d})\nonumber\\
        &=C_{V|U_{1},\ldots, U_{d}}(v|u_{1},\ldots,u_{d})\nonumber\\
        &=C_{V|U_{1},\ldots, U_{d}}(F_{Y}(y)|F_1(x_1),\ldots,F_d(x_d)).
        \label{eq: u-v-d-vine}
    \end{align}
Equation (\ref{eq: u-v-d-vine}), thus, allows us to express Equation (\ref{eq: d-vine-inv-con-dis}) as:
    \begin{equation}
        F^{-1}_{Y|X_1,\ldots,X_d}(\alpha|x_1,\ldots,x_d)=F^{-1}_Y\left(C^{-1}_{V|U_{1},\ldots, U_{d}}(\alpha|F_1(x_1),\ldots,F_d(x_d))\right).\\
        \label{eq: u-v-d-vine-inv-cond-dis}
    \end{equation}
    
This implies that the conditional quantile function can be expressed as the inverse of both the marginal distribution function of the response and the conditional copula quantile function given the observed \emph{PIT} value of $\boldsymbol{x}.$ Hence, we obtain an estimate of the conditional quantile function as:
    \begin{equation}
        \hat q_{\alpha}(x_1,\ldots,x_d)\coloneqq \hat F^{-1}_Y\left(\hat C^{-1}_{V|U_{1},\ldots, U_{d}}(\alpha|\hat u_{1},\ldots,\hat u_{d}%;\hat\thetab
        )\right),\\
        \label{eq: est-d-vine-cond-quan}
    \end{equation}
where $\hat u_{j} \coloneqq \hat F_j(x_j)$ is the estimated \emph{PIT} value of $x_j$, $j = 1, \ldots, d,$ and $\hat C^{-1}_{V|U_{1},\ldots, U_{d}}$ is an estimate of the conditional quantile function for $V$ given $U_1, \ldots, U_d$.

We now specify a D-vine copula to $(V,U_1,\ldots,U_d)^\top$ such that $V$ is fixed as the first node in the first tree. The order of the D-vine is $V$--$U_{k_1}$--$\ldots$--$U_{k_d}$, where $(k_1,\ldots,k_d)^\top$ is allowed to be an arbitrary permutation for $(1,\ldots,d)^\top$. This allows for flexibility because the order of the $U_j$ is chosen to maximize the conditional likelihood \citep{KRAUS20171}. Note that since we expressed $C_{V|U_{1},\ldots, U_{d}}(v| u_{1},\ldots, u_{d})$ in terms of nested $h$-functions in Equation (\ref{eq: joe97}), we can express the conditional quantile function in terms of inverse $h$-functions.

It is important to mention that $C^{-1}_{V|U_{1},\ldots, U_{d}}(\alpha|u_{1},\ldots,u_{d})$ is monotonically increasing in $\alpha$, meaning that crossing of quantile functions is not possible. As a result, the D-vine copula based quantile regression overcomes the pitfall of the benchmark \emph{lqr}.

\subsection{Estimation of \textit{D}-vine quantile regression} \label{subsec: est.-d-vine-reg}
To estimate the conditional quantile function in Equation (\ref{eq: est-d-vine-cond-quan}) utilizing Equation (\ref{eq: u-v-d-vine-inv-cond-dis}), we need to estimate the marginal distribution functions of all variables, including the response, and the conditional copula distribution. We proceed in a two step approach. First, we estimate the required marginal distribution functions and, secondly, estimate the $h$-functions based on the pair copulas. 

\subsubsection*{Marginal distribution function estimation} \label{subsubsec: marginal dist. est.}
In order to fit marginal distributions to our variables, we utilize parametric and a mixture of parametric univariate distributions. Fitting univariate parametric distributions to continuous data is a standard task in statistics. However, choosing an appropriate probability distribution for the univariate random variable needs a careful examination. We use maximum likelihood (\emph{ML}) estimation to estimate the parameters of the chosen distribution. 

Sometimes standard univariate parametric distributions do not fit the data at hand. A mixture of univariate parametric distributions enhances modeling flexibility considerably. Here we focus on a mixture of univariate normal distributions, which we express as a weighted sum of $S$ normal distributions $\{\psi_{1}(x;\boldsymbol{\Theta}_{1}), \ldots, \psi_{S}(x;\boldsymbol{\Theta}_{S})\}$. The weights $\{\omega_1, \ldots, \omega_S\}$ sum to one and each distribution has its own parameters $\boldsymbol{\Theta}$. We express this as 
    \begin{align}
        \begin{split}
            \text{minimize} \qquad&
            f(x; \boldsymbol{\Theta}_{1}, \ldots, \boldsymbol{\Theta}_{S} ) = \sum_{s = 1}^{S} \omega_{s} \psi_s(x;\boldsymbol{\Theta}_{s}),  \hspace{5 mm} \text{subject to}  \hspace{2 mm} \sum_{s=1}^{S}\omega_{s} = 1,
        \end{split}
        \label{eq: mini-marg-dis}
    \end{align}
where each normal distribution with density $\psi_s(x;\boldsymbol{\Theta}_{s})$ has its own parameters,  $\boldsymbol{\Theta}_{s} = (\mu_s, \sigma_s)^\top,$ $s = 1, \ldots, S.$ The distribution parameters are estimated using an expectation-maximization (EM) algorithm. The algorithm maximizes the conditional expected (complete-data) log likelihood at each M-step \citep{mclachlan2000finite}. 
%\emph{i.e.}
%$$\psi_s(x;\mu_s, \sigma_s) = \frac{1}{\sigma_s\sqrt{2\pi}} 
%  \exp\left( -\frac{1}{2}\left(\frac{x-\mu_s}{\sigma_s}\right)^{\!2}\,\right),$$
%where $\mu_s$ and $\sigma_s$ are the mean and standard deviation. Parameters 

After obtaining the estimated marginal distribution functions, $\hat{F}_Y$ and $\hat{F}_j$, $j = 1, \ldots, d,$ we transform the observed data to pseudo copula data $\hat{v}_i \coloneqq \hat{F}_{Y}(y_i)$ and $\hat{u}_{ij} \coloneqq \hat{F}_j(x_{ij})$, $j = 1, \ldots,d,$ $i = 1, \ldots, n.$ Let $\hat{\boldsymbol{u}}_i = (\hat{u}_{i1}, \ldots, \hat{u}_{id})^\top$ for $i = 1, \ldots, n,$ then $\{(\hat{v}_i, \hat{\boldsymbol{u}}_{i}): i = 1, \ldots, n\}$ is an approximate independent and identically distributed (\emph{i.i.d.}) sample of the \emph{PIT} random vector $(V, U_1, \ldots, U_d)^\top$. 

\subsubsection{Parametric \textit{D}-vine copula estimation} \label{subsubsec: par-d-vine-est}
In Equation (\ref{eq: est-d-vine-cond-quan}), we fit a D-vine with order $V-U_{k_1}-\ldots-U_{k_d}$ to pseudo copula data. The ordering $\boldsymbol{k} = (k_1, \ldots, k_d)^\top$ of the variables is chosen in a stepwise fashion to maximize the conditional log-likelihood (\emph{cll}) of the fitted D-vine. We define the \emph{cll} of an estimated D-vine with ordering $\boldsymbol{k}$, fitted parametric pair-copula families $\hat{\boldsymbol{\mathcal{F}}}$ and corresponding copula parameters $\hat{\boldsymbol{\theta}}$ as follows for given pseudo copula data $(\hat{\boldsymbol{v}} = (\hat{v}_1, \ldots, \hat{v}_2)^\top, \hat{\mathcal{U}} = \{\hat{\boldsymbol{u}}_i, i = 1, \ldots, n\})$:
    \begin{equation}
        cll(\boldsymbol{k}, \hat{\boldsymbol{\mathcal{F}}}, \hat{\boldsymbol{\theta}}; \hat{\boldsymbol{v}}, \hat{\mathcal{U}}) \coloneqq \sum_{i = 1}^{n} \text{log } c_{V|\boldsymbol{U}}(\hat{v}_{i}|\hat{\boldsymbol{u}}_{i}; \boldsymbol{k}, \hat{\boldsymbol{\mathcal{F}}}, \hat{\boldsymbol{\theta}}),
        \label{eq: cll}
    \end{equation}
where the conditional copula density $c_{V|\boldsymbol{U}}$ is expressed as:
    \begin{align*} 
            c_{V|\boldsymbol{U}}(\hat{v}_{i}|\hat{\boldsymbol{u}}_{i};\boldsymbol{k}, \hat{\boldsymbol{\mathcal{F}}}, \hat{\boldsymbol{\theta}}) =& 
            c_{VU_{k_1}}(\hat{v}_{i},\hat{u}_{{ik}_1};\hat{\boldsymbol{\mathcal{F}}}_{VU_{k_1}},\hat{\theta}_{VU_{k_1}})\times{}  &&\\
            \prod_{j=2}^{d} c_{VU_{k_j};U_{k_1},\ldots,U_{k_{j-1}}} &\Big(\hat{C}_{V|U_{k_1},\ldots,U_{k_{j-1}}}\big(\hat{v}_i|\hat{u}_{{ik}_1},\ldots,\hat{u}_{{ik}_{j-1}}\big), &&\\ \hat{C}_{U_{k_j}|U_{k_1},\ldots,U_{k_{j-1}}}&\big(\hat{u}_{{ik}_j}|\hat{u}_{{ik}_1},\ldots,\hat{u}_{{ik}_{j-1}}\big);\hat{\boldsymbol{\mathcal{F}}}_{VU_{k_j};U_{k_1},\ldots,U_{k_{j-1}}},\hat\theta_{VU_{k_j};U_{k_1},\ldots,U_{k_{j-1}}}\Big).&&
    \end{align*}

Here $\hat{\boldsymbol{\mathcal{F}}}_{VU_{k_j};U_{k_1},\ldots,U_{k_{j-1}}}$ and $\hat\theta_{VU_{k_j};U_{k_1},\ldots,U_{k_{j-1}}}$ denote the fitted pair copula family and its copula parameter estimate of $C_{VU_{k_{j}}; U_{k_1, \ldots, U_{k_{j-1}}}}$, respectively. \cite{KRAUS20171} provide more details on the D-vine regression forward selection algorithm. The algorithm sequentially constructs a D-vine while maximizing the model's \emph{cll} in each step and stops when the model's conditional log-likelihood can not be improved. This results in parsimonious models where only influential covariates are included. To account for model complexity, two standard selection criteria, which includes the number of needed parameters in the D-vine regression, are considered: AIC-corrected conditional log-likelihood ($cll^{AIC}$) and BIC-corrected conditional log-likelihood ($cll^{BIC}).$ The $cll^{AIC}$ is defined as:
    \begin{equation}
        cll^{AIC}(\boldsymbol{k}, \hat{\boldsymbol{\mathcal{F}}}, \hat{\boldsymbol{\mathcal{\theta}}}; \hat{\boldsymbol{v}}, \hat{\mathcal{U}}) \coloneqq
        -2 cll(\boldsymbol{k}, \hat{\boldsymbol{\mathcal{F}}}, \hat{\boldsymbol{\mathcal{\theta}}}; \hat{\boldsymbol{v}}, \hat{\mathcal{U}}) + 2 |\hat{\boldsymbol{\theta}}|,
        \label{eq: cll-caic}
    \end{equation} and the $cll^{BIC}$ is defined as
    \begin{equation}
        cll^{BIC}(\boldsymbol{k}, \hat{\boldsymbol{\mathcal{F}}}, \hat{\boldsymbol{\mathcal{\theta}}}; \hat{\boldsymbol{v}}, \hat{\mathcal{U}}) \coloneqq
        -2 cll(\boldsymbol{k}, \hat{\boldsymbol{\mathcal{F}}}, \hat{\boldsymbol{\mathcal{\theta}}};     \hat{\boldsymbol{v}}, \hat{\mathcal{U}}) + log(n) |\hat{\boldsymbol{\theta}}|.
        \label{eq: cll-cbic}
    \end{equation}
\cite{KRAUS20171} state that the AIC-corrected $cll^{AIC}$ is sufficient since only up to two parameters per copula family are estimated. Example \ref{ex: for-d-vine-algo} illustrates the forward selection algorithm on a four-dimensional vector $(Y, X_1, X_2, X_3)^\top$.

\begin{example}[Forward selection algorithm on four-dimensional vector]\label{ex: for-d-vine-algo}
Suppose we have a four-dimensional vector $(Y, X_1, X_2, X_3)^\top$, where $Y$ denotes the response variable and $(X_1, X_2, X_3)^\top$ are the covariates. After transforming the data to pseudo copula data $(\hat{v}_i, \hat{u}_{i1}, \hat{u}_{i2}, \hat{u}_{i3})^\top$ for $i = 1, \ldots, n$, we choose the pair $(V, U_j)^\top$, $j = 1, 2, 3,$ which has the largest $cll$ among the D-vines $V-U_j$, $j = 1, 2, 3$. Assume the pair $(V, U_3)^\top$ has the largest conditional log-likelihood, then we have two remaining variables: $U_1$ and $U_2$. We investigate if adding $U_1$ or $U_2$ would improve the $cll$ of the model. Assuming that adding $U_1$ is better than adding $U_2$ and if it also improves over $V-U_3$, then we update the D-vine to $V-U_3-U_1$ from $V-U_3.$ Next, we study if adding the remaining variable $U_2$ would improve the conditional log-likelihood of $V-U_3-U_1$ or not. If, for example, the model's $cll$ with order $V-U_3-U_1$ is equal to the model's $cll$ with order $V-U_3-U_1-U_2$, the algorithm returns the D-vine with order $V-U_3-U_1.$
\end{example}

\subsection{Estimation of critical event probabilities using the \textit{D}-vine regression approach} \label{subsec: est-critical-prob}

For risk assessment, we are interested in estimating the probability of a critical event $Y > c $  for a fixed $c$ under the conditions $\bm{X} = \bm{x},$ \emph{i.e.} 
    \begin{equation}
        \begin{aligned}
            \alpha_{c}(\bm{x})  \coloneqq P(Y > c | \bm{X} = \bm{x}) & =  1 - P(Y \leq c |\bm{X} = \bm{x}) \\
            & = 1 - F_{Y|X_1,\ldots,X_d}(c|x_1,\ldots,x_d).
        \end{aligned}
    \label{eq: crit-alpha}
    \end{equation} 
In view of Equation (\ref{eq: lqr cond. quan.}) and Equation (\ref{eq: est-d-vine-cond-quan}), we see $\alpha_{c}(\bm{x})$ is a function of the inverse of $q_{\alpha}(\bm{x})$ with respect to $\alpha.$ The conditional distribution function in Equation (\ref{eq: crit-alpha}) can be estimated using a bisection algorithm to invert the conditional quantile function in the \emph{lqr} case. However, Equation (\ref{eq: crit-alpha}) is obtained directly via the Rosenblatt transform \citep{Rosenblatt} in the case for D-vine copula based approach.

The bisection algorithm is a root-finding algorithm applied to the function with respect to $\alpha$ in Equation (\ref{eq: lqr cond. quan.}). The algorithm selects a sub-interval in which the root must lie in, and the sub-interval is iteratively narrowed down until the solution is reached. The bisection algorithm is necessary for the \emph{lqr} since the \emph{lqr} does not have a joint density associated. See Algorithm \ref{alg: bisection algorithm} in Appendix \ref{Append: Algorithms} for pseudo code.

On the other hand, the Rosenblatt transform for D-vines is easily available for an observation and is implemented in the R-package \textbf{rvinecopulib} \citep{rvinecopulib}. The Rosenblatt transform will be utilized to compute the conditional distribution function at the chosen threshold \emph{c}. For a random vector $\boldsymbol{V} = (V_1, \ldots, V_p)^\top$ with joint distribution function $F$, let the conditional distribution function of $V_j$ given $V_1, \ldots, V_{j-1}$ denoted by $F_{j|1, \ldots, j-1},$ $j = 2, \ldots, p.$ Then, the random vector $\boldsymbol{U}$, where
 $U_1 \coloneqq F(V_1)$ and $U_j \coloneqq F_{j|1, \ldots, j-1}(V_j|V_1, \ldots, V_{j-1})$ for $j = 2, \ldots, p,$ is joint uniformly distributed. With the use of Equation (\ref{eq: d-vine-density-indices}), we can express $F_{j|1, \ldots, j-1}(v_j|v_1, \ldots, v_{j-1})$ for observed values $V_j = v_j, \ldots, V_1 = v_1,$ as $F_{Y|X_1, \ldots, X_d}(c|x_1, \ldots, x_d)$ for arbitrary values of $c$ given the value $\bm{X} = \bm{x}.$

\section{Data analysis}\label{sec: data analysis}
\begin{figure}[h!]
		\centering
		\includegraphics[width=1\textwidth]{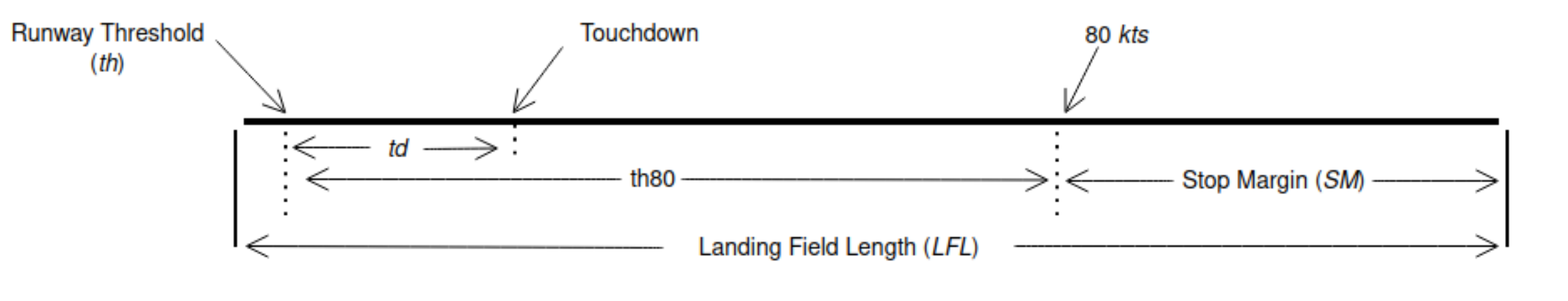}
		\caption{Runway with markings such as \emph{th}, \emph{td}, \emph{th80} and \emph{SM}.}
		\label{fig: runwayThresh}
	\end{figure}

The critical event probabilities are estimated as in Equation (\ref{eq: crit-alpha}) for the response \emph{th80} given the contributing factors $\bm{X}$ in Table \ref{tab: Contr. Fact.}. Mathematically, we write
    \begin{equation}
        \alpha_{c}(\bm{x}_i) = 1- P(th80_i \leq c  |\bm{X}_i = \bm{x}_i)
        \label{eq: risk-alpha}
    \end{equation}
for $i = 1, \ldots, 711,$ and an appropriately chosen threshold $c$. The choice of \emph{c} must be less than the Landing Field Length (\emph{LFL}), leaving an appropriate \emph{SM} distance for the aircraft to exit the runway or completely stop. Figure \ref{fig: runwayThresh} illustrates a runway with \emph{th80} as well as the distance from runway threshold to touchdown, denoted by (\emph{td}). It is important to mention that we include \emph{td} as a contributing factor. Next, more details are provided on the estimation of the associated D-vine regression model. 

\newpage
\subsection{\textit{D}-vine regression model estimation} \label{subsec: d-vine-model}

Discussed in  Section \ref{subsec: est.-d-vine-reg}, we first estimate the marginal distribution functions of all variables parametrically or via a mixture of univariate normal distributions. Figure \ref{fig: density-hist-vari} in Appendix \ref{AppB} displays, on the left, fitted univariate and a mixture of univariate normal density estimates on the original scale (raw data) for all variables and their respective histograms on the pseudo copula scale, \emph{PIT} values, on the right of each sub-figure. In addition, Table \ref{tab: fitted-par-dist} in Appendix \ref{AppB} lists the univariate distribution family fitted for each variable. Some of the fitted parametric distributions consists of more than two parameters, \emph{i.e.} the skew Student t, which is utilized for fitting \emph{refAP} and \emph{hws}.
    
\begin{figure}[h!]
		\centering
		\includegraphics[width=1\textwidth]{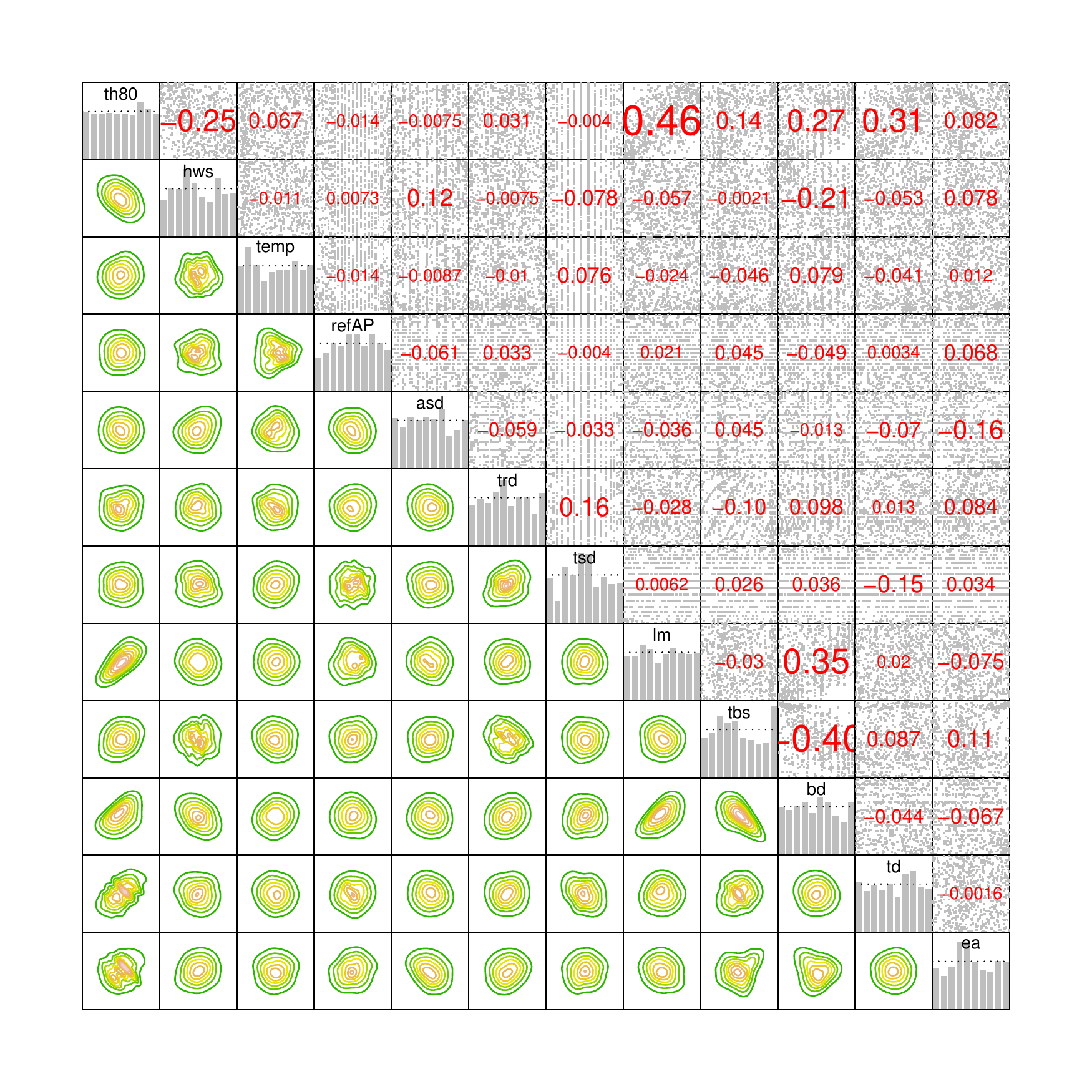}
		\caption{Dependence exploration for flight data (normalized contour plots on the lower diagonal panels, histograms of the \emph{PIT} values on the diagonal panels and pairwise scatter plots of the \emph{PIT} values with an associated empirical Kendall's $\tau$ value on the upper diagonal panels).}
		\label{fig: contour}
	\end{figure}

Following the fitted marginal distribution functions, we display in Figure \ref{fig: contour} three different panels: the normalized contour plots on the lower diagonal, the \emph{PIT} of the variables on the copula scale and the Kendall's tau dependence between the variables on the upper diagonal. The normalized contour plots represent the transformation of a bivariate copula density to a bivariate distribution with standard normal margins and density $g(z_1, z_2),$ where 
\begin{equation*}
g(z_1, z_2) = c(\Phi(z_1), \Phi(z_2)) \cdot \phi(z_1) \cdot \phi(z_2).
\end{equation*}

Here $\Phi(\cdot)$ and $\phi(\cdot)$ are the distribution and density of a standard normal distribution, respectively. Note that the dependence between \emph{th80} and \emph{lm} is the strongest with an empirical $\hat{\tau}_{Kendall} = 0.46$, while \emph{hws} is negatively dependent on \emph{th80} with $\hat\tau_{Kendall} = -0.25$. The pseudo copula data for each variable, which is derived from the fitted marginal distribution, is approximately uniform as seen by the diagonal panels of Figure \ref{fig: contour}. Among the 711 flights we are studying, we observe only 33 unique values for $tsd$. This is apparent in the pairwise scatter plots in which vertical and horizontal lines are visible in the upper diagonal panels of Figure \ref{fig: contour}. Further, the normalized contour plots on the lower diagonal panels assess departure from the Gaussian copula assumption if the contour lines are not elliptical. This is seen, for example, in the contour panel between $th80$ and the contributing factor $lm$.

We fit a D-vine copula based regression using the R package \textbf{vinereg} \citep{vinereg}. To allow for easy simulation, we utilize only parametric distributions for the pair copula families. Table \ref{tab: d-vine-summary} gives a summary of the estimated conditional quantile function: 
    \begin{equation}
        \hat q_{\alpha}(hws,\ldots,ea)=\hat F^{-1}_{th80}\left(\hat C^{-1}_{V_{th80}|U_{hws},\ldots, U_{ea}}(\alpha|\hat u_{hws},\ldots,\hat u_{ea}%;\hat\thetab
        )\right),\\
        \label{eq: d-vine-resp-cov-est}
    \end{equation}
where $th80$ is the response and the covariates ($hws, \ldots, ea$) are the contributing factors listed in Table \ref{tab: Contr. Fact.}.
    
    \begin{table}[!htbp]
    \centering
        \caption{Summary output of fitted $D$-vine quantile regression for the response variable $th80$ given the contributing factors. The summary table includes the following: fitted pair copula family $\hat{\boldsymbol{\mathcal{F}}}_{1k_j;k_1,\ldots,k_{j-1}}$, its parameters $\hat{\boldsymbol{\theta}}_{1k_j;k_1,\ldots,k_{j-1}}$, the estimated Kendall's tau $\hat{\tau}_{1k_j;k_1,\ldots,k_{j-1}}$, the log-likelihood $ll_{1k_j;k_1,\ldots,k_{j-1}}$, the $ll_{1k_j;k_1,\ldots,k_{j-1}}^{aic}$, the $ll_{1k_j;k_1,\ldots,k_{j-1}}^{bic}$ and a \texttt{p\_value}.}
        \scalebox{0.65}{
        \begin{tabular}{|rrrrrrrrrr|}
        \hline
        name & $k_j$ & $k_1, \ldots, k_{j-1}$ & $\hat{\boldsymbol{\mathcal{F}}}_{1k_j;k_1,\ldots,k_{j-1}}$ & $\hat{\boldsymbol{\theta}}_{1k_j;k_1,\ldots,k_{j-1}}$ & $\hat{\tau}_{1k_j;k_1,\ldots,k_{j-1}}$ & $ll_{1k_j;k_1,\ldots,k_{j-1}}$ & $ll_{1k_j;k_1,\ldots,k_{j-1}}^{aic}$ & $ll_{1k_j;k_1,\ldots,k_{j-1}}^{bic}$ & p\_value\\ 
        \hline
        lm & 7 & - & bb8 & 3.62, 0.84 & 0.45 & 196.19  & -388.37 & -379.24 & $< 0.00$\\ 
        td & 10 & 7 & frank & 4.43 & 0.42 & 151.77 & -301.54 & -296.97 & $< 0.00$\\ 
        hws & 2 & 10, 7 & gaussian & -0.50 & -0.34 & 96.11 & -190.22 & -185.65 & $< 0.00$\\ 
        ea & 11 & 2, 10, 7 & gaussian & 0.44 & 0.29  & 75.85 & -149.70  & -145.13 & $< 0.00$ \\ 
        asd & 5 & 11, 2, 10, 7 & gaussian & 0.45 & 0.30 & 81.00 & -160.00 & -155.43 & $< 0.00$ \\ 
        temp & 3 & 5, 11, 2, 10, 7 & frank & 1.86 & 0.20  & 34.87 & -67.74 & -63.18 & $< 0.00$ \\ 
        tbs & 8 & 3, 5, 11, 2, 10, 7 & frank & 2.11  & 0.22 & 42.56 & -83.13 & -78.56 & $< 0.00$ \\ 
        bd & 9 & 8, 3, 5, 11, 2, 10, 7 & frank & 2.02 & 0.22 & 36.25 & -70.49 & -65.93 & $< 0.00$ \\ 
        trd & 6 & 9, 8, 3, 5, 11, 2, 10, 7 & gumbel & 1.09 & 0.09 & 10.55 & -19.11 & -14.54 &  $< 0.00$ \\ 
        refAP & 4 & 6, 9, 8, 3, 5, 11, 2, 10, 7 & gumbel & 1.12 & -0.11 & 9.11  & -16.22  & -11.66 &  $< 0.00$\\ 
        \hline
        \end{tabular}}
        \label{tab: d-vine-summary}
    \end{table}

 The variables are labeled with indices described in Table \ref{tab: labels}. Here $k_1$ denotes the first contributing factor added, and $k_j$, likewise, denotes the $j^{\text{th}}$ contributing factor added in the D-vine. Table \ref{tab: d-vine-summary} includes the fitted pair copula family $\hat{\boldsymbol{\mathcal{F}}}_{1k_j;k_1,\ldots,k_{j-1}}$, its parameters $\hat{\boldsymbol{\theta}}_{1k_j;k_1,\ldots,k_{j-1}}$, the estimated Kendall's tau $\hat{\tau}_{1k_j;k_1,\ldots,k_{j-1}}$, the log-likelihood $ll_{1k_j;k_1,\ldots,k_{j-1}}$, the $ll_{1k_j;k_1,\ldots,k_{j-1}}^{aic}$ in Equation (\ref{eq: cll-caic}) and the $ll_{1k_j;k_1,\ldots,k_{j-1}}^{bic}$ in Equation (\ref{eq: cll-cbic}). It also includes the \texttt{p\_value} of a likelihood ratio test. The test compares the D-vine regression model with $k_1, \ldots, k_{j-1}$ contributing factors to another one with an additional factor $k_j$ using conditional log likelihoods. 
 
 We write the log likelihood for the pair copula term $c_{1k_j;k_1, \ldots, k_{j-1}}$ as 
 \begin{equation*}
        \begin{split} 
            ll_{1k_{j}; k_1, \ldots, k_{j-1}} & \coloneqq \sum_{i = 1}^{n} ln ( c_{1k_{j};k_1, \ldots, k_{j-1}}(C_{1|k_1, \ldots, k_{j-1}}(\hat{u}_{i1}|\hat{u}_{ik_{1}}, \ldots, \hat{u}_{ik_{j-1}}),\\
            &C_{k_j|k_1, \ldots, k_{j-1}}(\hat{u}_{ik_{j}}|\hat{u}_{ik_{1}}, \ldots, \hat{u}_{ik_{j-1}}))).
        \end{split}
    \end{equation*}
 Since the conditional log likelihood of the model $U_1|U_{k_1}, \ldots, U_{k_j}$ is given by
 \begin{equation*}
    cll_{1|k_1, \ldots, k_j} = \sum_{m = 1}^{j} ll_{1k_{m}; k_1, \ldots, k_{m-1}},
 \end{equation*}
 we have that the likelihood ratio test statistic is
 \begin{equation*}
     2 \times \left[ cll_{1|k_1, \ldots, k_{j}} - cll_{1|k_1, \ldots, k_{j-1}}\right] = 2 \times ll_{1k_{j}: k_1, \ldots, k_{j-1}}.
 \end{equation*}
 If the corresponding \texttt{p\_value} $< 0.05,$ we have significant evidence that the additional contributing factor $X_{k_{j}}$ is improving the model fit over the model with $X_{k_{1}}, \ldots, X_{k_{j-1}}$ factors.
 
     \begin{table}[!htbp]
     \centering
         % title of Table
         % used for centering table
        \scalebox{1}{
        \begin{threeparttable}
            \caption{Variable labels in the summary table of the estimated $D$-vine quantile regression.}
            \label{tab: labels} % is used to refer this table in the text
        \begin{tabular}{|c| c c c c c c c c c c c|} % centered columns (4 columns)
        \hline %inserts double horizontal lines
        \textbf{Variable} & \emph{th80} & \emph{hws} & \emph{temp} & \emph{refAP} &  \emph{asd} & \emph{trd}  & \emph{lm}  &  \emph{tbs} & \emph{bd}  & \emph{td} & \emph{ea}  \\ [0.5ex] % inserts table
        %heading
        \hline % inserts single horizontal line
        \textbf{Label} & 1 & 2 & 3 & 4 & 5  &  6  & 7 & 8 & 9 & 10 & 11\\
        \hline %inserts single line
        \end{tabular}
        \end{threeparttable}}
        
    \end{table}
 
 One of the contributing factors, \emph{tsd}, was not selected as a candidate to maximize the $cll$. The contributing factor \emph{tsd} is also the variable with a few unique observations among the total $711$ observations. For more details, Table \ref{tab: vine-trees}  in Appendix \ref{AppB} gives a full summary of the fitted D-vine copula.

\subsection{\textit{lqr} model estimation} \label{subsec: lqr-model-est}
\begin{figure}[H]
		\centering
		\includegraphics[width=.47\textwidth]{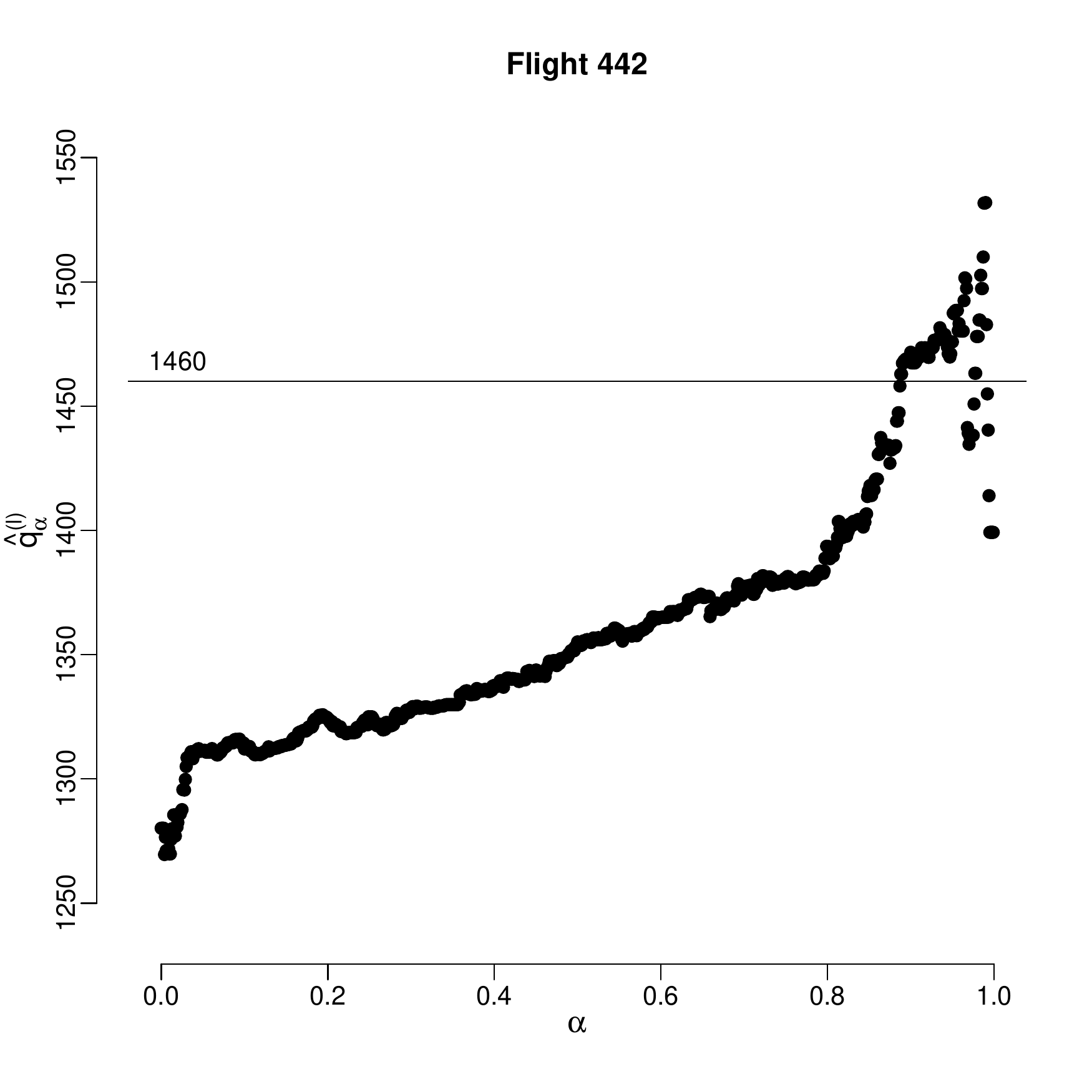}
		\caption{Non unique quantile levels, $\alpha$, at the same threshold $c = 1460$ for flight $442$.}
		\label{fig: flight 442}
	\end{figure}

Using the R package \textbf{quantreg} \citep{quantregP}, we fit a \emph{lqr} model for the response variable, \emph{th80}, conditioned on all contributing factors listed in Table \ref{tab: Contr. Fact.}. We write this as:
    \begin{equation}
        q_{\alpha}^{(l)}(hws,\ldots,ea):=F^{-1}_{th80|\bm{X}}(\alpha|hws,\ldots,ea)
        \label{eq: lqr_quant-fit-est}
    \end{equation}
for different quantile levels $\alpha \in (0,1)$. Table \ref{tab: lqr-fit-0.5-0.9} gives a summary of the estimated \emph{lqr} at two quantile levels: $\alpha = 0.5$ and $\alpha = 0.9$. The significance of the contributing factors on the response changes for different quantile levels. For instance, at $\alpha = 0.5,$ the contributing factor \emph{bd} has \texttt{p\_value} $< 0.001,$ whereas, at $\alpha = 0.9,$ the \texttt{p\_value} increases to more than $ 0.80.$ The method used to calculate standard error estimates is the bootstrap as proposed by \cite{koenker2001quantile}.
\begin{table}[H]
    \centering
    \scalebox{0.96}{
    \begin{threeparttable}
     \caption {Two fitted \emph{lqr} models at $\alpha = 0.5$ and $ \alpha = 0.9$ with different significance of contributing factors.} 
    \label{tab: lqr-fit-0.5-0.9}   
    \begin{tabular}{|rrrrrrrr|}
        \toprule
        Variable (quantile)& \multicolumn{2}{c}{$th80$ ($\alpha = 0.5$)} & & & \multicolumn{2}{c}{$th80$ ($\alpha = 0.9$)}   &   \\ 
        \cline{2-4} \cline{6-8} 
            &  Value   & Std. Error    & \texttt{p\_value}    &                                                      
	        &  Value   & Std. Error    & \texttt{p\_value}                                              \\    \hline
            (Intercept) & 362.91 & 512.81 & &  & $-$733.18 & 1,327.30 &  \\
            hws & $-$32.13 & 3.91 & $^{***}$&  & $-$40.86 & 5.11 & $^{***}$\\ 
            temp & 4.01 & 0.74 & $^{***}$ & &  3.27 & 1.18 &  $^{***}$\\ 
            refAP & $-$1.55 & 0.44 & $^{***}$ & & $-$0.29 & 1.31 & \\ 
            asd & 27.59 & 3.54 & $^{***}$& & 38.22 & 5.28 & $^{***}$\\ 
            trd & 8.99 & 4.19 & $^{**}$ & & 16.32 & 10.16 & \\ 
            tsd & 14.21 & 14.55 & & & 13.95 & 29.66 & \\ 
            lm & 3.95 & 0.42 & $^{***}$ & & 5.86 & 0.47 & $^{***}$\\ 
            tbs & 25.47 & 5.25& $^{***}$ & & 10.02 & 3.59 & $^{***}$\\ 
            bd & 21.19 & 6.12 & $^{***}$ & &  0.88 & 4.30 & \\ 
            td & 1.01 & 0.03 & $^{***}$ & & 0.89 & 0.08 & $^{***}$\\ 
            ea & 212.14 & 37.42 & $^{***}$ & & 209.27 & 55.53 & $^{***}$\\
            \hline \\[-1.8ex] 
            Observations & 711  & & & & & &\\ 
            \hline 
            \hline \\[-1.8ex] 
            \textit{Note:}  & \multicolumn{2}{r}{$^{*}$p$<$0.1; $^{**}$p$<$0.05; $^{***}$p$<$0.01} & & & & & \\ 
            \bottomrule
        \end{tabular}
        \end{threeparttable}}
    \end{table}
Note that some estimates are not unique with respect to $\alpha.$ For example, it is possible to have two or more different quantile levels with the same estimate $\hat{q}_{\alpha}^{(l)}$ for one observation. Figure \ref{fig: flight 442} displays two $\alpha$ values with the same conditional quantile estimate $\hat{q}_{\alpha}^{(l)} = 1460$ for flight 442. This illustrates the problem of quantile crossing, which occurs in our data. 

\subsection{Risky flight probability estimation} \label{subsec: high-risk-prob-est}
 The number of flights having an estimate of $\alpha_c(\bm{x}_i)$ greater than $10^{-13}$ for $i = 1, \ldots, 711$ is listed in Table \ref{tab: nonzero-est} for three different thresholds $c$ in meters, $c = (2200, 2400, 2500)^\top.$ The estimated critical event probability, $\hat{\alpha}_c(\bm{x}_i)$, was obtained via the bisection algorithm for the estimated \emph{lqr} of (\ref{eq: lqr_quant-fit-est}) and via the Rosenblatt transform of (\ref{eq: d-vine-resp-cov-est}). The estimated \emph{lqr} model uses all contributing factors for the probability estimation, no variable selection was performed. On the other hand, the estimated D-vine regression model uses ten of the contributing factors, leaving \emph{tsd} out. The three different thresholds were chosen according to the landing field length described earlier in Section \ref{sec: data descri.}. 
 
 The estimated \emph{lqr} model was not able to provide estimates greater than $10^{-13}$ beyond the observed maximum distance of \emph{th80} (2,606.77 \emph{m}). The estimated $D$-vine regression model, however, provided nonzero estimates beyond the observed maximum distance of \emph{th80} for over $50\%$ of the flights. For the D-vine approach, we further consider two cases where only Gaussian pair copula families are used (D-vine$^{Gauss}$) and several parametric pair copula families are used (D-vine$^{Par}$).

	\begin{table}[H] \centering 
	
	    \scalebox{1}{
	    \begin{threeparttable}
	    \caption{Number of flights having estimates of $\alpha_c(\bm{x}_i)$ larger than $10^{-13}$ from three different models $lqr$, D-vine$^{Gauss}$ and D-vine$^{Par}$ for three different thresholds $c$ in \emph{m}.}
	    \label{tab: nonzero-est} 
        \begin{tabular}{@{\extracolsep{5pt}} ccccccc} 
            \\[-1.8ex]\hline 
            \hline \\[-1.8ex] 
            & \textbf{\emph{lqr} (\%)} &  &\textbf{D-vine$^{Gauss}$}(\%)  & &\textbf{D-vine$^{Par}$}(\%) &    \\ 
            \hline \\[-1.8ex] 
            \multicolumn{1}{c|}{\textbf{$2200$ \emph{m}}} & $206$ (28.97\%)& & 709 (99.72\%)& & $708$ (99.58\%) &  \\ 
            \hline \\[-1.8ex] 
            \multicolumn{1}{c|}{\textbf{$2400$ \emph{m}}} & $59$ (8.30\%) &  & 709 (99.72\%)& & $703$ (98.87\%) &    \\ 
            \hline \\[-1.8ex] 
            \multicolumn{1}{c|}{\textbf{$2500$ \emph{m}}} & $23$ (3.23\%)&  & 706 (99.30\%)& & $691$ (97.19\%) &    \\ 
            \hline \\[-1.8ex] 
            \end{tabular}
            \end{threeparttable}}
        
    \end{table} 
    
\subsubsection*{Identification of risky flights} \label{subsubsec: high-risk-flight-ident}
 We investigate flights which have an estimated risk probability $\hat{\alpha}_c(\bm{x}_i) > 0.001$ and threshold $c = 2500$ \emph{m} for $i = 1, \ldots, 711,$. Among the 711 flights, we identify 41 such flights using the D-vine$^{Par}$ approach. The largest risk probability estimate within this group is $0.202$. In comparison to the other two approaches, the D-vine$^{Gauss}$ identifies the least. Table \ref{tab: comparison-tab} lists the five identified risky flights using the D-vine$^{Gauss}$ approach, which are also among those identified in the \emph{lqr} and the D-vine$^{Par}$ approaches. The estimated risk probabilities from each approach are included in Table \ref{tab: comparison-tab}. 
 
\begin{table}[h!]
\centering
\begin{threeparttable}

\caption{List of the five identified flights having an estimated risk probability $> 10^{-3}$ using the D-vine$^{Gauss}$ approach compared to the other approaches with their respective probabilities.}
\label{tab: comparison-tab}
\begin{tabular}{rrrr}
  \hline
  flight & lqr\_est\_prb & D\_vine\_Gauss & D\_vine\_par \\ 
  \hline
    1 & 0.007 & 0.540 & 0.202 \\ 
    2 & 0.004 & 0.368 & 0.013 \\ 
    3 & 0.007 & 0.015 & 0.054 \\ 
    4 & 0.004 & 0.010 & 0.036 \\ 
    5 & 0.006 & 0.014 & 0.077 \\ 
   \hline
\end{tabular}
\end{threeparttable}
\end{table}
 
\begin{figure}[h!]
		\centering
		\includegraphics[width= 14.25cm, height = 14.5cm]{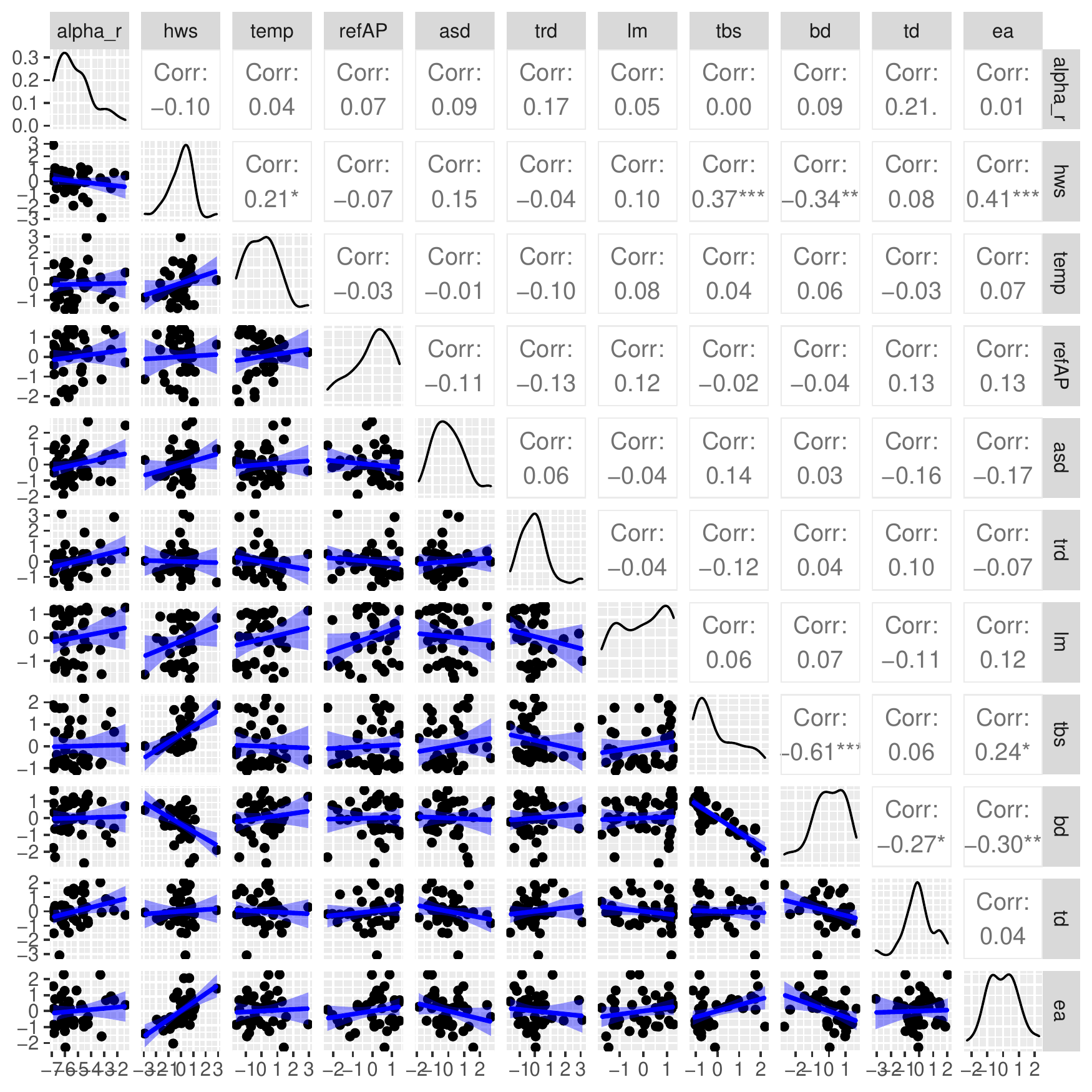}
		\caption{Pairwise scatter plots displayed on the lower diagonal panels, pairwise kendall's tau dependence on the upper diagonal and density plots of the variables on the diagonal panels.}
		\label{fig: pairwise-lin}
	\end{figure}

 Because we are interested in studying the relationship between the contributing factors and the estimated risk probabilities for the risky flights, we will focus on the 41 flights identified using the D-vine$^{Par}$ approach. The $41$ estimated risk probabilities fall in the range $(0.001, 0.203)$, therefore, we transform the estimates to the real number line $(-\infty, \infty)$ using the logit function:
    \begin{equation}
        \eta_{r} = \text{logit}(\hat{\alpha}_{c}(\bm{x}_r)) = ln\left( \frac{\hat{\alpha}_{c}(\bm{x}_r)}{1-\hat{\alpha}_{c}(\bm{x}_r)}\right) 
        \label{eq: logit-scale}
    \end{equation}
for $r = 1, \ldots, 41.$ Since, also, we want to compare the chosen contributing factors resulted from the D-vine$^{Par}$ approach to each other in terms of how much impact they have on the estimated risk probability $\eta_r$, we standardize the factors $X_{k_{j}}$, $j = 1, \ldots, 10$, using 
\begin{equation*}
    z_{rk_{j}} \coloneqq \frac{x_{rk_{j}} - \bar{x}_{k_{j}}}{\sqrt{\frac{\sum_{r = 1}^{41} (x_{rk_{j}} - \bar{x}_{k_{j}})^{2}}{N-1}}}, \hspace{2mm} \text{with} \hspace{2mm} \bar{x}_{k_{j}} = \frac{1}{41}\sum_{r = 1}^{41} x_{rk_{j}}
\end{equation*}
 for $r = 1, \ldots, 41.$ Figure \ref{fig: pairwise-lin} shows pairwise scatter plot matrices on the lower diagonal and pairwise empirical Kendall's $\tau$ on the upper diagonal. On the diagonal panels, we see density plots of the estimated risk probabilities on the logit scale and the contributing factors on the standardized scale. We add in blue fitted linear regression lines for each variable paired with another with $90\%$ pointwise confidence intervals. For example, \emph{hws} and \emph{tbs} have a positive linear relationship shown by an upward pointing blue line in the scatter plot and an empirical Kendall's $\tau$ dependence equal to $0.37.$    

\begin{figure}[H]
		\centering
		\includegraphics[width=14.2cm, height=9.5cm]{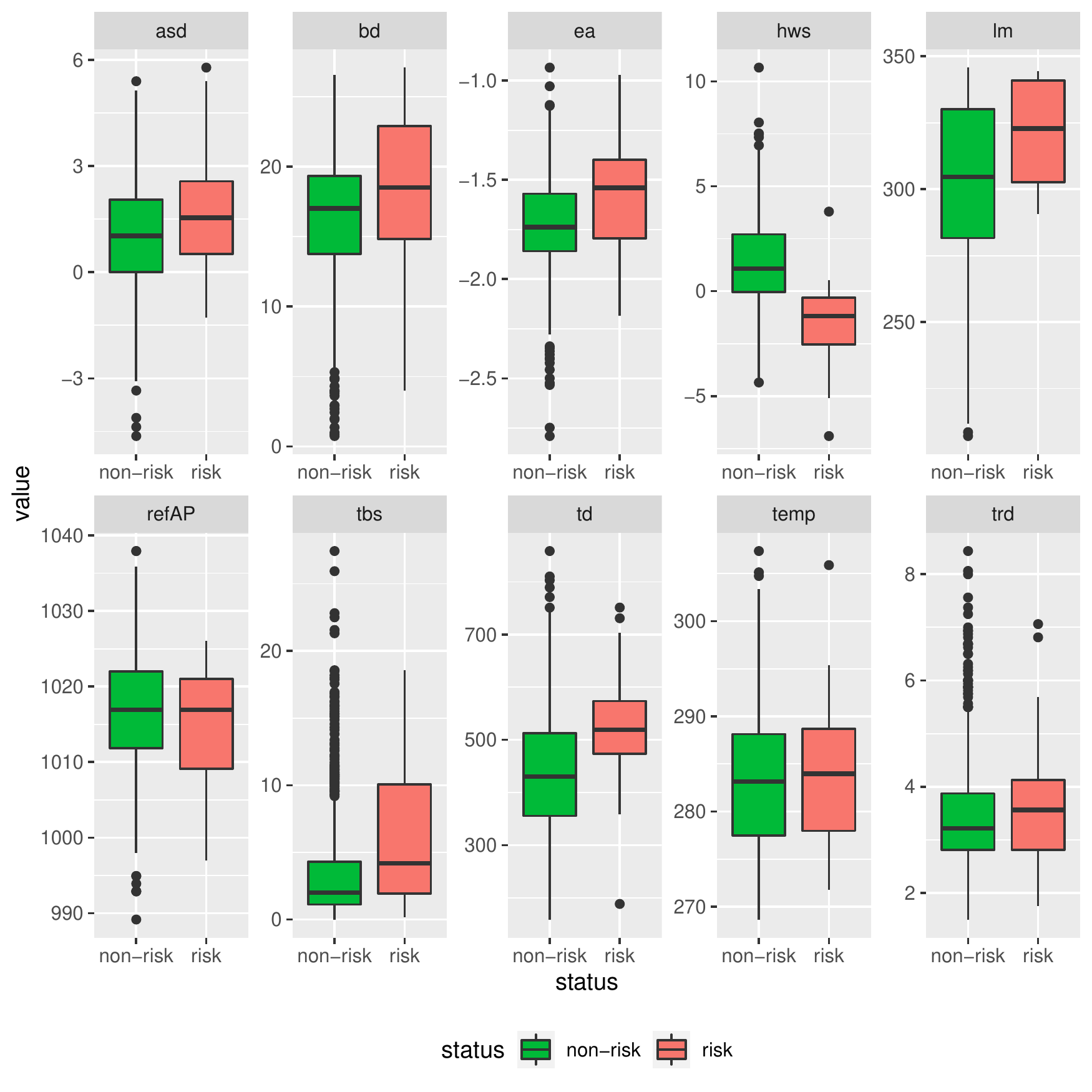}
		\caption{Box-plots of each contributing factor indicating the two different flight groups, red: risk and green: non-risk.}
		\label{fig: box-plots-risk-n}
	\end{figure}

\newpage
Based on the scatter plots in Figure \ref{fig: pairwise-lin}, we fit a multiple linear regression as the relationship between the estimated logit of risk probabilities and the standardized contributing factors could be explained linearly. We estimate the following multiple regression equation:
\begin{equation}
    \alpha_{c}(\bm{x}_{rk_{j}}) = \beta_0 + \sum_{j = 1}^{10}\beta_{j} z_{rk_{j}} + \epsilon_r,
    \label{eq: lnr-eq-reg-risk}
\end{equation}
where $\alpha_{c}(\bm{x}_{rk_{j}})$ denotes the estimated logit of the risk probability for flight $r$, $r = 1, \ldots, 41,$ and $\epsilon_r$ denotes normally distributed error with zero mean and $\sigma^2$ variance for flight $r.$ Table \ref{tab: summary-out-lnr-reg} gives the estimated summary output of Equation (\ref{eq: lnr-eq-reg-risk}). The output shows that a one standard deviation increase in \emph{hws} leads to a $-1.44$ decrease in estimated logit of the risk probabilities. This estimated model gives an adjusted $R^2 = 0.80$, which states that $80\%$ of the variability in the estimated logit of the risk probabilities is accounted for by the ten D-vine selected contributing factors. Figure \ref{fig: box-plots-risk-n} shows two types of box-plots for each contributing factor (red: risk and green: non-risk). The risk group corresponds to flights with an estimated risk probability $>10^{-3}$, while the non-risk group corresponds to flights with an estimated risk probability $<10^{-3}$. We note some contributing factors, \emph{i.e.} \emph{hws}, differ in their observed values for the two flight groups.

\begin{table}[ht]
    \centering

    \begin{threeparttable}
    
    \caption{Summary output of the estimated multiple linear regression in (\ref{eq: lnr-eq-reg-risk}). The table includes estimated coefficients (Estimate), standard errors (Std. Error), t statistic values (t value) and \texttt{p\_values}.}
    
    \label{tab: summary-out-lnr-reg}
    
    \begin{tabular}{rrrrr}
        \hline
        & Estimate & Std. Error & t value & Pr($>$$|$t$|$) \\ 
        \hline
        (Intercept) & -5.20 & 0.10 & -53.82 & 0.00 \\ 
        hws & -1.44 & 0.17 & -8.62 & 0.00 \\ 
        ea & 1.15 & 0.15 & 7.78 & 0.00 \\ 
        td & 1.06 & 0.13 & 8.01 & 0.00 \\
        asd & 1.02 & 0.12 & 8.33 & 0.00 \\
        tbs & 0.96 & 0.24 & 3.97 & 0.00 \\
        bd & 0.83 & 0.26 & 3.19 & 0.00 \\
        lm & 0.47 & 0.12 & 4.02 & 0.00 \\
        trd & 0.46 & 0.10 & 4.38 & 0.00 \\
        temp & 0.28 & 0.11 & 2.48 & 0.02 \\ 
        refAP & -0.22 & 0.11 & -1.97 & 0.06\\ 
         
        \hline
    \end{tabular}
    \end{threeparttable}
\end{table}

\begin{figure}[H]
		\centering
		\includegraphics[width=0.65\textwidth]{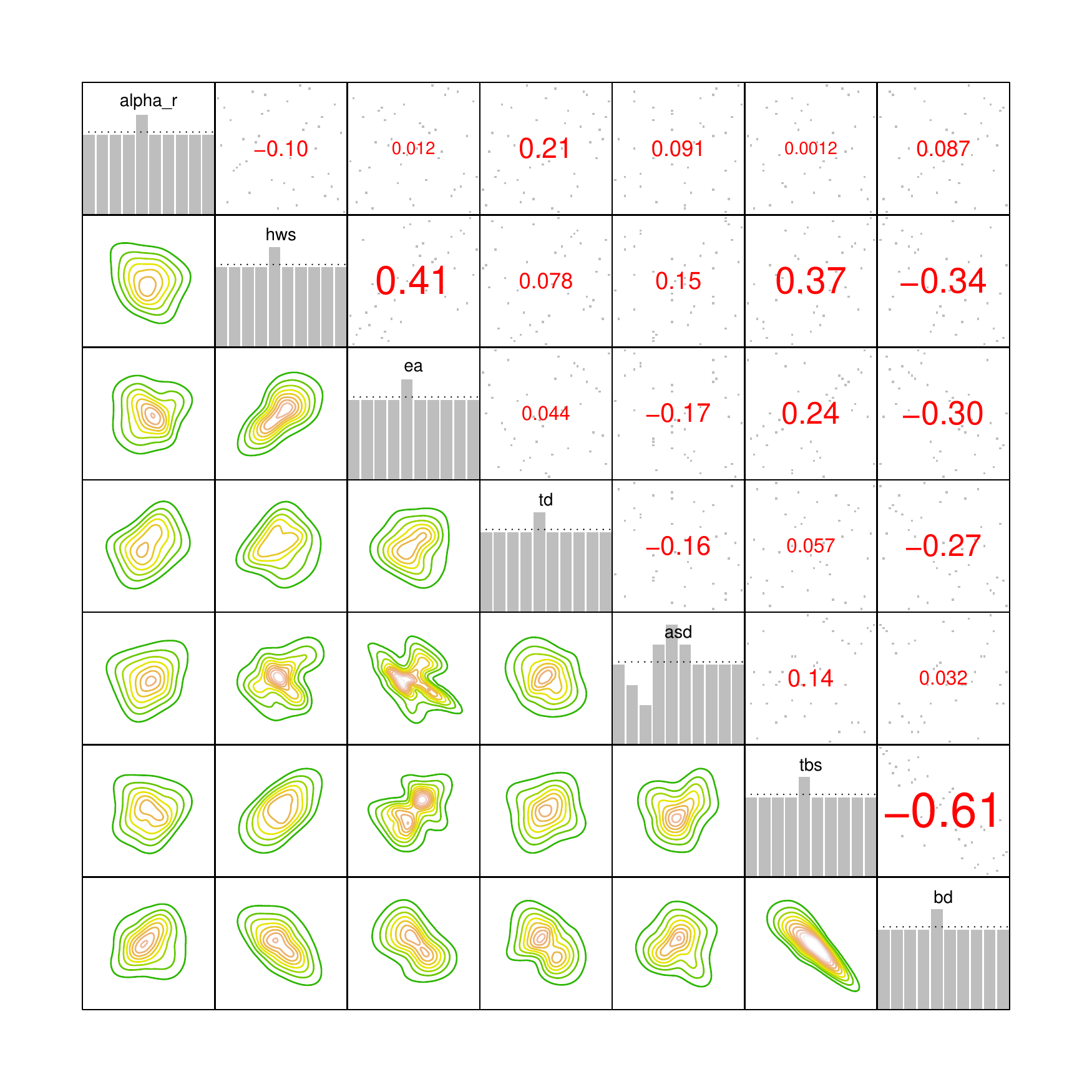}
		\caption{Dependence exploration for risky flights (normalized contour plots on the lower diagonal, histograms of the \emph{PIT} values on the diagonal and pairwise scatter plots of the \emph{PIT} values with associated empirical Kendall’s $\tau$ values on the upper diagonal panels).}
		\label{fig: ranked-depend-contour}
	\end{figure}

We rank the contributing factors based on their impact on the response variable measured by the size of the corresponding regression coefficient in Table \ref{tab: summary-out-lnr-reg}. For the highest six contributing factors (\emph{hws}, \emph{ea}, \emph{td}, \emph{asd}, \emph{tbs}, \emph{bd}), we fit another multiple linear regression. The summary output, given in Table \ref{tab: ranked-lnr-reg-summ}, gives an adjusted $R^2 = 0.60$, i.e. $60\%$ of the variance in the estimated logit of the risk probabilities can be explained by these factors. In addition, we study the pairwise dependence between these contributing factors. Figure \ref{fig: ranked-depend-contour} displays similar panels to those in Figure \ref{fig: contour} except that on the diagonal panels we fitted empirical distribution functions to the margins rather than univariate parametric distributions. We avoid parametric marginal distribution fitting due to the small sample size, $r = 41$. Among the lower diagonal panels, two panels show departure from the Gaussian copula assumption by the non elliptical shape of their contour lines. The contour panel between \emph{hws} and \emph{ea} shows a positive tail dependence, while the panel between \emph{tbs} and \emph{bd} shows a strong negative tail dependence. 

\begin{table}[H]
    \centering
    \begin{threeparttable}
    \caption{Summary output of a fitted multiple linear regression on a subset of the contributing factors.}
    \label{tab: ranked-lnr-reg-summ}
    
        \begin{tabular}{rrrrr}
        \hline
        & Estimate & Std. Error & t value & Pr($>$$|$t$|$) \\ 
        \hline
        (Intercept) & -5.20 & 0.14 & -38.42 & 0.00 \\ 
        bd & 1.12 & 0.32 & 3.50 & 0.00 \\
        hws & -1.09 & 0.21 & -5.24 & 0.00 \\ 
        tbs & 1.07 & 0.31 & 3.46 & 0.00 \\
        td & 1.05 & 0.18 & 5.95 & 0.00 \\
        ea & 1.02 & 0.20 & 5.11 & 0.00 \\ 
        asd & 0.96 & 0.17 & 5.75 & 0.00 \\ 
        \hline
    \end{tabular}
    \end{threeparttable}
\end{table}

\newpage
\section{Discussion} \label{sec: conclusion}    

We have shown that the probability of a flight decelerating to the controllable speed of 80 \emph{kts} exceeding a chosen threshold $c$ can be estimated conditioned on a set of influencing factors. The D-vine regression approach of \cite{KRAUS20171} was utilized since it is able to model very small conditional probabilities flexibly and estimate them analytically. In particular, we identified 41 flights, among 711, of the same aircraft type landed at the same airport having distance to controllable speed greater that 2500 meter with an estimated probability $>10^{-3}$. The marginal effect of each contributing factor for these risky flights was further studied and ranked from the highest as follows: brake duration, headwind speed, time brake started, touchdown, equivalent acceleration and approach speed deviation. Also, the joint pairwise behavior among the contributing factors for the risky flights was investigated. We show that there is a non symmetric dependence between time brake started and brake duration as well as between headwind speed and equivalent acceleration. 

Unlike the physics-based approach of \cite{Drees} and \cite{PCE}, the proposed D-vine approach does not require simulation nor subset simulation for small probabilities. Additionaly, the D-vine approach does not require the discretization of the continuously measured factors and does not require experts' knowledge in network design as needed in Bayesian network approaches. Using the D-vine approach, also, nonzero probability estimates beyond the observed maximum distance of the controllable speed were provided in contrast to the classical \emph{lqr} approach.  

We have shown that our statistical approach is capable of quantifying the risk probability of runway overrun (the probability of a flight exceeding a chosen threshold at the controllable speed of 80 \emph{kts}) and ranking the effects of the influencing factors on runway overrun. Similar scenarios, such as early landing and veer-offs, are also of a great importance in aviation safety and will be the focus of future investigations.

\begin{appendix}
\section{Algorithms} \label{Append: Algorithms}

\subsection{Bisection Algorithm} 
The algorithm used to estimate (\ref{eq: crit-alpha}) for the \emph{lqr} in (\ref{eq: lqr cond. quan.}) is given the following: 

\begin{algorithm} 
        \SetKwFunction{isOddNumber}{isOddNumber}

        \SetKwInOut{KwIn}{Input}
        \SetKwInOut{KwOut}{Output}

        \For{$i = 1, \ldots, n$}{
            Choose $a$ and $b$ $\in (0,1)$ with $a < b.$\\
            Evaluate $\hat{q}_{\alpha}(x_1, \ldots, x_d)$ in (\ref{eq: lqr cond. quan.}) for both $\alpha = a$ and $\alpha = b$.\\
            \eIf{$\hat{q}_{c}(x_1, \ldots, x_d) \in (\hat{q}_{a}(x_1, \ldots, x_d), \hat{q}_{b}(x_1, \ldots, x_d))$}{
                1. Increase $a$ by $\delta,$ $\delta \in (0,1).$\\
                2. Repeat.
            }{
            \eIf{$\hat{q}_{c}(x_1, \ldots, x_d) \neq \hat{q}_{a}(x_1, \ldots, x_d)$ $\&$ $\hat{q}_{c}(x_1, \ldots, x_d) \neq \hat{q}_{b}(x_1, \ldots, x_d)$}{
                1. Increase $b$ by $\delta$ and decrease $a$ by $\delta,$ $\delta \in (0,1).$\\
                2. Repeat.}{
            \eIf{$\hat{q}_{c}(x_1, \ldots, x_d) = \hat{q}_{a}(x_1, \ldots, x_d)$ $\&$ $\hat{q}_{c}(x_1, \ldots, x_d) \neq \hat{q}_{b}(x_1, \ldots, x_d)$}{Return $a$.}{Return $b$.}
            }
         }
    }

    \caption{Bisection algorithm for the \emph{lqr}}
    \label{alg: bisection algorithm}
    \end{algorithm}
    
\section{Figures and Tables} \label{AppB}

\subsection{Pairwise Scatter Plots and Dependence}
We display on the lower diagonal panels pairwise scatter plots, on the diagonal panels density plots and on the upper diagonal panels the empirical Kendall's $\tau$. In the lower diagonal panels, we also add fitted linear regression lines in blue for each variable together with $90\%$ confidence intervals.  

    \begin{figure}[H]
		\centering
		\includegraphics[width=1\textwidth]{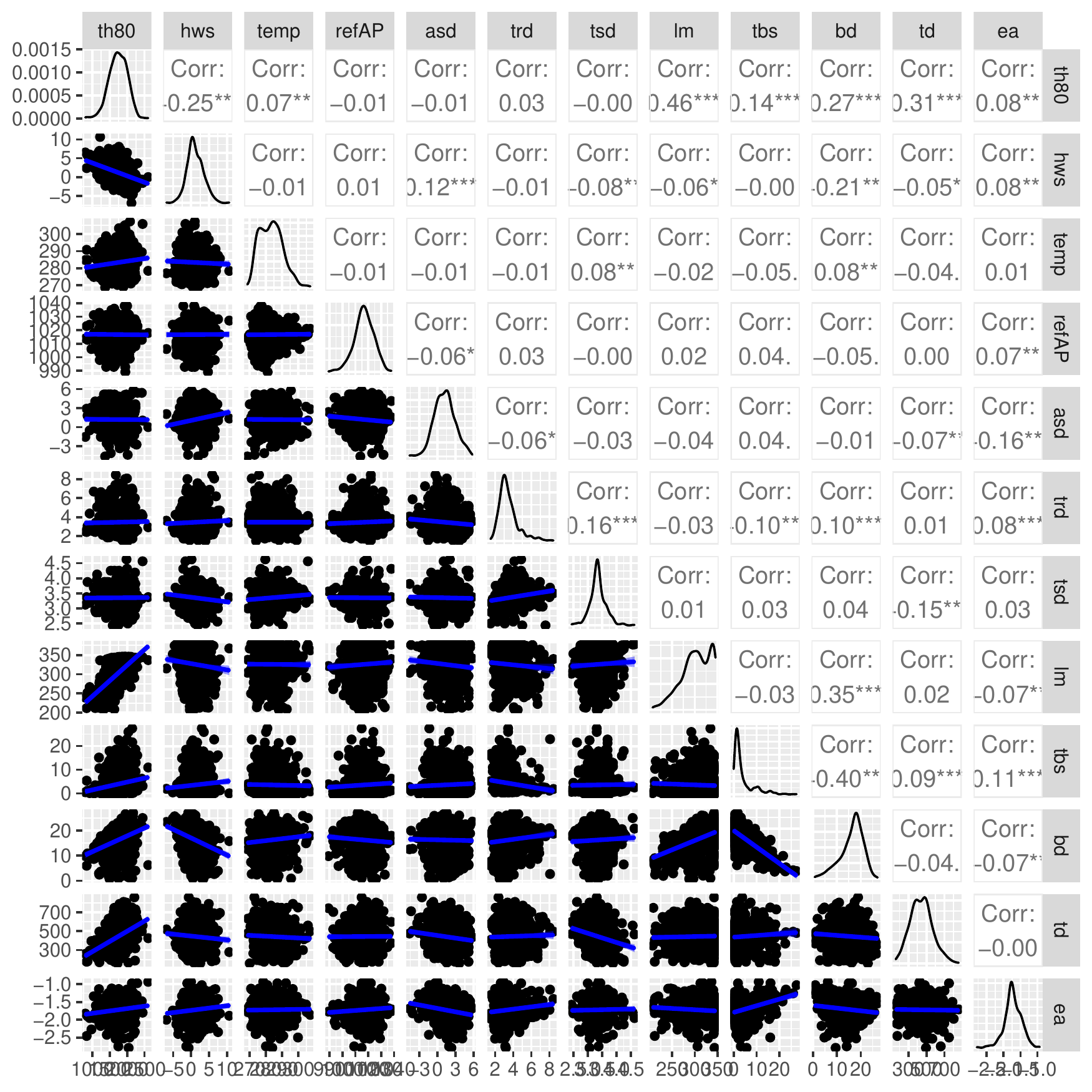}
		\caption{}
		\label{fig: pairwise-raw}
	\end{figure}

\subsection{Fitted Marginal distribution functions}
We list the fitted distribution families and their parameter estimates for all variables in our data set.

	\begin{table}[H] 
        \caption{Fitted marginal distributions for all variables and their parameter estimates.} 
        \label{tab: fitted-par-dist} 
        \centering
        \scalebox{1}{
        \begin{tabular}{@{\extracolsep{5pt}} ccc} 
            \\[-1.8ex]\hline 
            \hline \\[-1.8ex] 
            \textbf{Variable} & \textbf{\emph{Selected distribution} } &\textbf{\emph{Parameter estimates} (\%)}\\ 
            \hline \\[-1.8ex] 
            \multicolumn{1}{c|}{\textbf{th80}} & Normal & [$\hat\mu, \hat\sigma$] = [1739.943, 259.2278] \\ 
            \hline \\[-1.8ex] 
            \multicolumn{1}{c|}{\textbf{hws}} &   Skew Student t. & [$\hat\xi, \hat\omega, \hat\alpha, \hat\nu$] = [-0.7578, 2.9865, 1.4194, 19] \\ 
            \hline \\[-1.8ex] 
            \multicolumn{1}{c|}{\textbf{temp}} &  Log-Normal & [$\hat\mu, \hat\sigma$] = [5.6462, 0.0245] \\ 
            \hline \\[-1.8ex] 
            \multicolumn{1}{c|}{\textbf{refAP}} & Skew Student t. & [$\hat\xi, \hat\omega, \hat\alpha, \hat\nu$] = [1023.067, 9.8146, -1.3456, 9] \\
            \hline \\[-1.8ex] 
            \multicolumn{1}{c|}{\textbf{asd}} &  Skew Normal & [$\hat\xi, \hat\omega, \hat\alpha$] = [0.3789, 1.8669, 0.6545] \\ 
            \hline \\[-1.8ex] 
            \multicolumn{1}{c|}{\textbf{trd}} &  Generalized Extreme Value & [$\hat\mu, \hat\sigma, \hat\nu$] = [2.9832, 0.7539, 0.0580] \\ 
            \hline \\[-1.8ex] 
            \multicolumn{1}{c|}{\textbf{tsd}} &  Log Normal &  [$\hat\mu, \hat\sigma$] = [1.2064, 0.0826]\\ 
            \hline \\[-1.8ex] 
            &   & [$\hat\mu_1, \hat\mu_2, \hat\mu_3, \hat\mu_4$] = [265.3788, 304.4632, 336.5114, 342.8597]\\
            \\[-1.8ex] 
            \multicolumn{1}{c|}{\textbf{lm}} & Mixture of Normals & [$\hat\sigma_1, \hat\sigma_2, \hat\sigma_3, \hat\sigma_4$] = [24.928, 16.6916,  4.0202,  1.4208] \\
            \\[-1.8ex] 
            & & [$\hat\omega_1, \hat\omega_2, \hat\omega_3, \hat\omega_4$] = [0.2636, 0.4957, 0.1013, 0.1393]\\
            \hline \\[-1.8ex] 
            \multicolumn{1}{c|}{\textbf{tbs}} &  Generalized Extreme Value & [$\hat\mu, \hat\sigma, \hat\nu$] = [1.6125, 1.5624,  0.5771] \\ 
            \hline \\[-1.8ex] 
            &   & [$\hat\mu_1, \hat\mu_2$] = [10.7978, 18.3855]\\
            \\[-1.8ex] 
            \multicolumn{1}{c|}{\textbf{bd}} & Mixture of Normals & [$\hat\sigma_1, \hat\sigma_2$] = [4.3706, 2.8982] \\
            \\[-1.8ex] 
            & & [$\hat\omega_1, \hat\omega_2$] = [0.282 0.7180]\\
            \hline \\[-1.8ex] 
            \multicolumn{1}{c|}{\textbf{td}} &  Gamma & [$\hat\alpha, \hat\beta$] = [12.6204, 0.0285] \\ 
            \hline \\[-1.8ex] 
            \multicolumn{1}{c|}{\textbf{ea}} &  Skew Normal & [$\hat\xi, \hat\omega, \hat\alpha$] = [-1.722, 0.2500, 0.9385]\\ 
            \hline \\[-1.8ex]
            \end{tabular}}
    \end{table} 
    
\subsection{Histograms of Variables on the Original Scale and Pseudo Copula Scale}
These histograms display the univariate and a mixture of univariate normal density estimates on the original scale as well as on the pseudo copula scale for all variables in our data.     
    \begin{figure}
        \begin{subfigure}{.46\textwidth}
            \centering
            % include first image
            \includegraphics[width=1\linewidth]{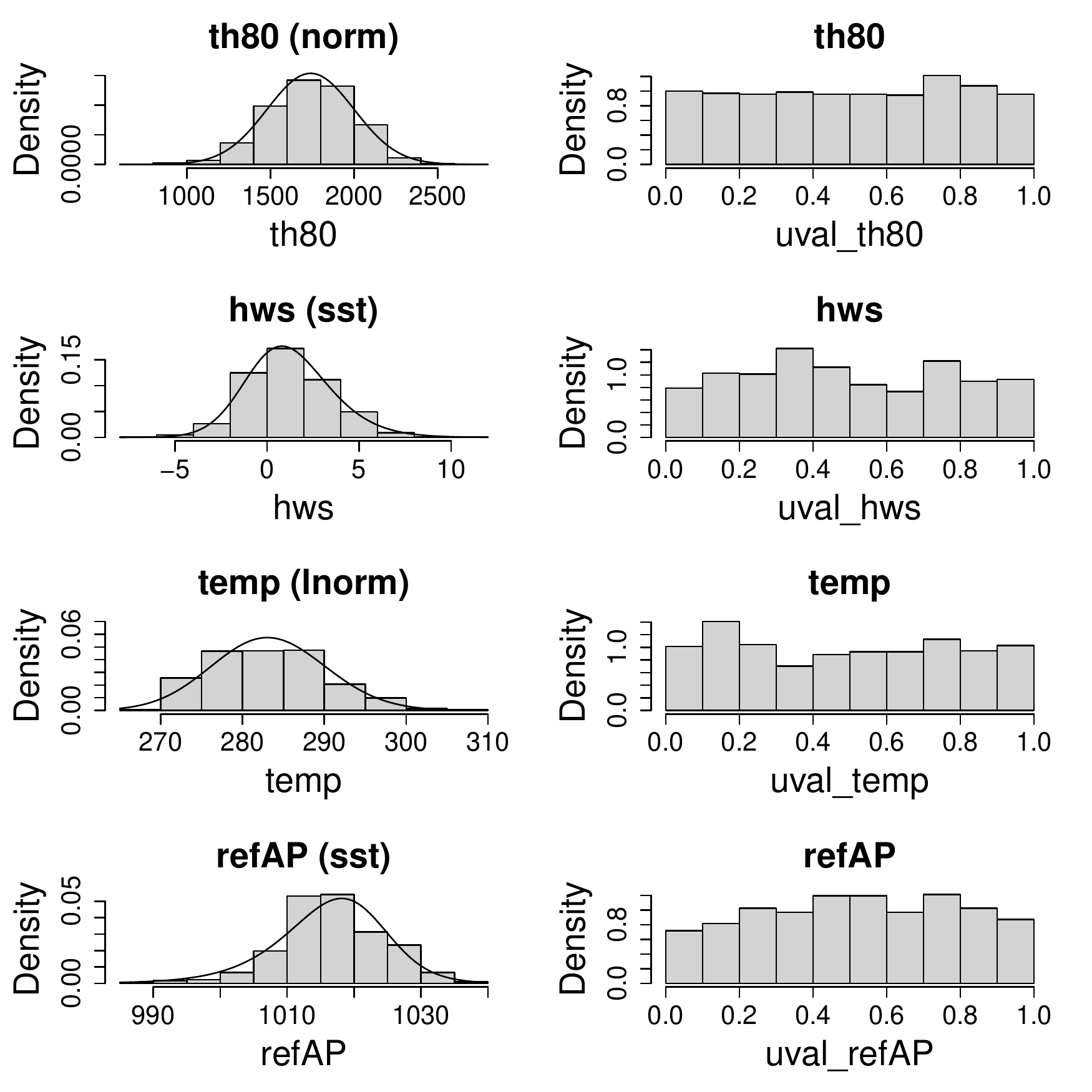}  
            \caption{Density estimates and histograms for: \emph{th80}, \emph{hws}, \emph{temp} \& \emph{refAP}.}
            \label{fig: sub-first}
        \end{subfigure}
        \begin{subfigure}{.46\textwidth}
            \centering
            % include second image
            \includegraphics[width=1\linewidth]{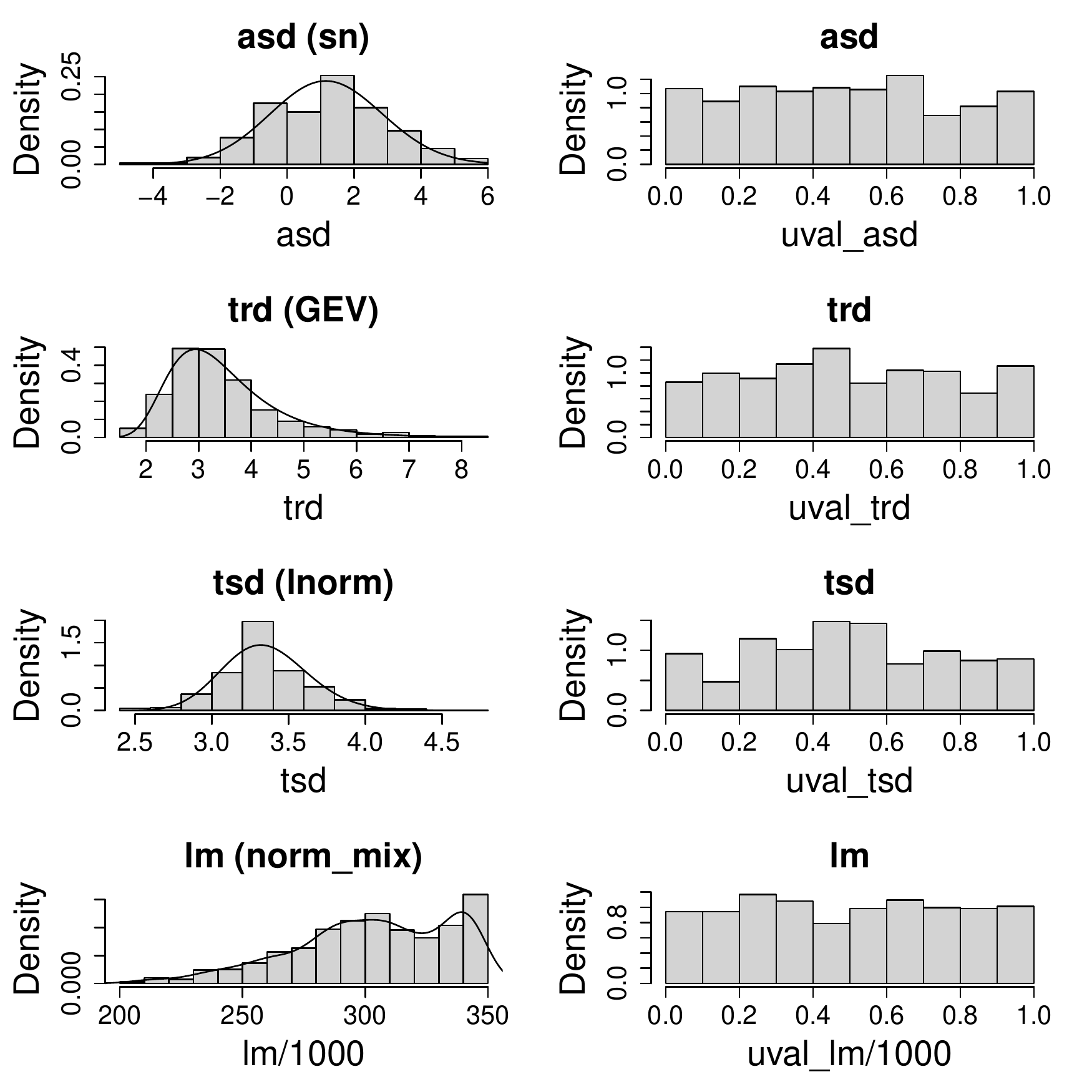}  
            \caption{Density estimates and histograms for: \emph{asd}, \emph{trd}, \emph{tsd} \& \emph{lm}.}
            \label{fig: sub-second}
        \end{subfigure}
        \newline
        \begin{subfigure}{.46\textwidth}
            \centering
            % include third image
            \includegraphics[width=1\linewidth]{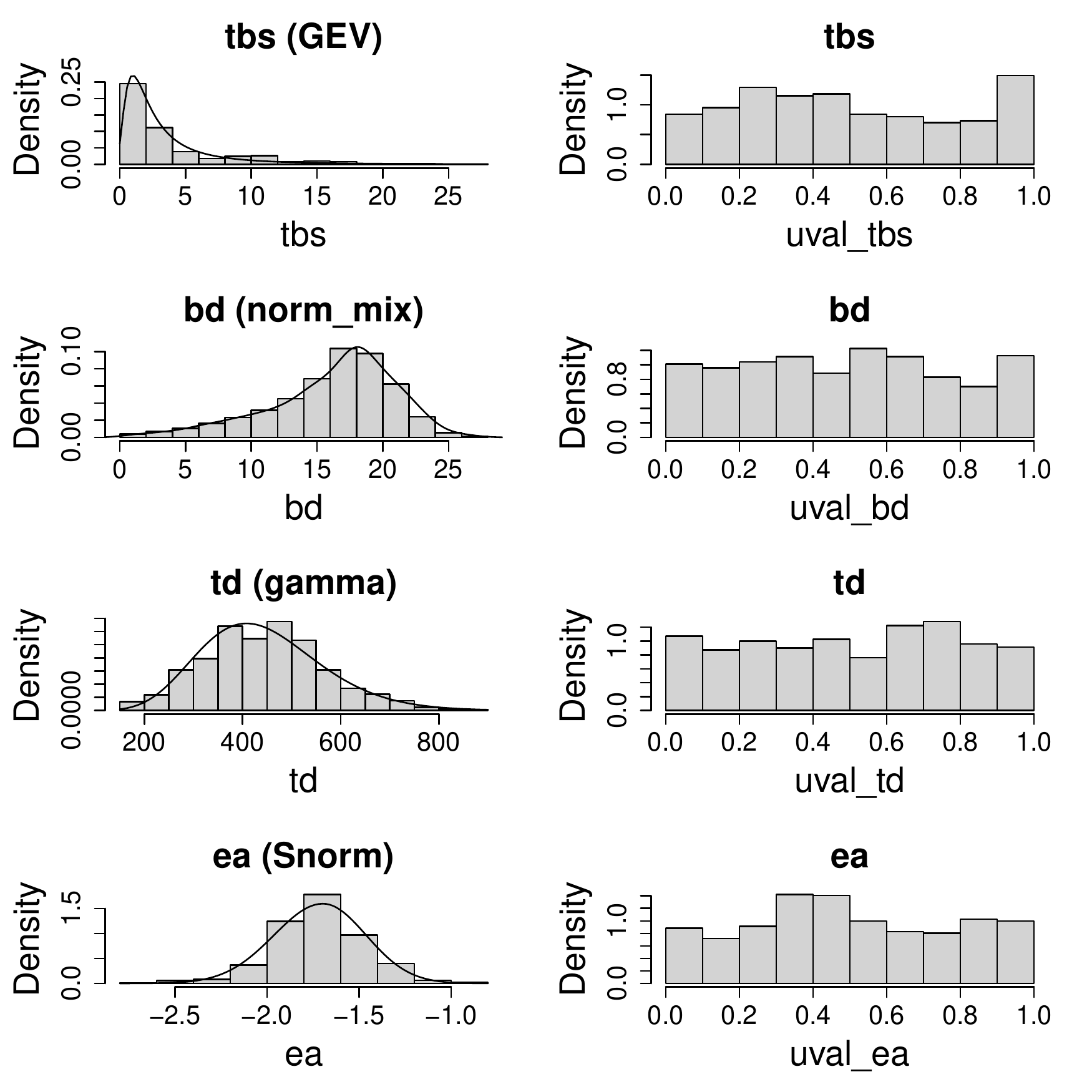}  
            \caption{Density estimates and histograms for: \emph{tbs}, \emph{bd}, \emph{td} \& \emph{ea}.}
            \label{fig: sub-third}
        \end{subfigure}
        \caption{Density estimates on the original scale for all variables on the left of each sub figure and their corresponding histograms on the copula scale.}
        \label{fig: density-hist-vari}
    \end{figure}
    
\newpage
    
\subsection{Fitted D-vine Copula}    
We give the full summary of the fitted D-vine copula, which consists of ten trees. The table includes entities such as the fitted pair copula family, its parameters, degrees of freedom (df), fitted Kendall's tau, the \emph{cll} value as well as the upper and lower tail dependence coefficients (utd, ltd). %Note that the fits among the contributing factors are needed to determine the conditional response distribution. 
    \begin{table}[H] 
        \centering
        \scalebox{0.8}{
        \begin{tabular}{rrrlllrlrrrrr}
            \toprule
            tree  & conditioned & conditioning  & family & rotation & parameters & df & tau & loglik & utd & ltd\\
            \midrule
            \rowcolor{green}
            1  & 1, 7 &   & bb8 & 180 & 3.62, 0.84 & 2 & 0.45 & 196.19 & 0.00 & 0.00 \\
            1  & 7, 10 &  & joe & 180 & 1.03 & 1 & 0.02 & 1.16 & 0.00 & 0.04\\
            1  & 10, 2 &  & gaussian & 0 & -0.09 & 1 & -0.06 & 2.70 & 0.00 & 0.00\\
            1  & 2, 11 &  & bb1 & 0 & 0.11, 1.03 & 2 & 0.07 & 6.48 & 0.04 & 0.00\\
            1  & 11, 5 &  & frank & 0 & -1.47 & 1 & -0.16 & 18.96 & 0.00 & 0.00\\
            1  & 5, 3 &   & indep & 0 &  & 0 & 0.00 & 0.00 & 0.00 & 0.00 \\
            1  & 3, 8 &   & frank & 0 & -0.37 & 1 & -0.04 & 1.44 & 0.00 & 0.00\\
            1  & 8, 9 &   & clayton & 90 & 1.22 & 1 & -0.38 & 175.76 & 0.00 & 0.00\\
            1  & 9, 6 &   & frank & 0 & 0.90 & 1 & 0.10 & 7.58 & 0.00 & 0.00\\
            1  & 6, 4 &   & indep & 0 &  & 0 & 0.00 & 0.00 & 0.00 & 0.00\\ 
            \addlinespace
            \rowcolor{green}
            2  & 1, 10 & 7& frank & 0 & 4.43 & 1 & 0.42 & 151.77 & 0.00 & 0.00\\
            2  & 7, 2 & 10& gaussian & 0 & -0.11 & 1 & -0.07 & 3.79 & 0.00 & 0.00\\
            2  & 10, 11 & 2 & indep & 0 &  & 0 & 0.00 & 0.00 & 0.00 & 0.00\\
            2  & 2, 5 & 11  & bb8 & 0 & 1.71, 0.80 & 2 & 0.14 & 18.92 & 0.00 & 0.00\\
            2  & 11, 3 & 5  & indep & 0 &  & 0 & 0.00 & 0.00 & 0.00 & 0.00\\
            2  & 5, 8 & 3 & gumbel & 0 & 1.05 & 1 & 0.05 & 2.31 & 0.06 & 0.00\\
            2  & 3, 9 & 8  & clayton & 180 & 0.16 & 1 & 0.07 & 7.36 & 0.01 & 0.00\\
            2  & 8, 6 & 9  & gaussian & 0 & -0.09 & 1 & -0.06 & 3.06 & 0.00 & 0.00\\
            2  & 9, 4 & 6  & clayton & 270 & 0.13 & 1 & -0.06 & 4.41 & 0.00 & 0.00\\
            \addlinespace
            \rowcolor{green}
            3  & 1, 2 & 10, 7 & gaussian & 0 & -0.50 & 1 & -0.34 & 96.11 & 0.00 & 0.00\\
            3  & 7, 11 & 2, 10  & bb8 & 270 & 1.25, 0.91 & 2 & -0.08 & 5.87 & 0.00 & 0.00\\
            3  & 10, 5 & 11, 2 & clayton & 90 & 0.15 & 1 & -0.07 & 5.85 & 0.00 & 0.00\\
            3  & 2, 3 & 5, 11  & indep & 0 &  & 0 & 0.00 & 0.00 & 0.00 & 0.00\\
            3  & 11, 8 & 3, 5  & joe & 0 & 1.39 & 1 & 0.18 & 52.06 & 0.35 & 0.00\\
            3  & 5, 9 & 8, 3  & indep & 0 &  & 0 & 0.00 & 0.00 & 0.00 & 0.00\\
            3  & 3, 6 & 9, 8  & indep & 0 &  & 0 & 0.00 & 0.00 & 0.00 & 0.00\\
            3  & 8, 4 & 6, 9  & clayton & 0 & 0.06 & 1 & 0.03 & 1.14 & 0.00 & 0.00\\
            \addlinespace
            \rowcolor{green}
            4  & 1, 11 & 2, 10, 7  & gaussian & 0 & 0.44 & 1 & 0.29 & 75.85 & 0.00 & 0.00\\
            4  & 7, 5 & 11, 2, 10  & clayton & 270 & 0.11 & 1 & -0.05 & 3.60 & 0.00 & 0.00\\
            4  & 10, 3 & 5, 11, 2  & frank & 0 & -0.35 & 1 & -0.04 & 1.26 & 0.00 & 0.00\\
            4  & 2, 8 & 3, 5, 11  & indep & 0 &  & 0 & 0.00 & 0.00 & 0.00 & 0.00\\
            4  & 11, 9 & 8, 3, 5  & t & 0 & 0.09, 11.68 & 2 & 0.05 & 6.16 & 0.01 & 0.01\\
            4  & 5, 6 & 9, 8, 3  & clayton & 270 & 0.09 & 1 & -0.04 & 2.54 & 0.00 & 0.00\\
            4  & 3, 4 & 6, 9, 8  & joe & 90 & 1.11 & 1 & -0.06 & 3.45 & 0.00 & 0.00\\
            \addlinespace
            \rowcolor{green}
            5  & 1, 5 & 11, 2, 10, 7  & gaussian & 0 & 0.45 & 1 & 0.30 & 81.00 & 0.00 & 0.00\\
            5  & 7, 3 & 5, 11, 2, 10  & joe & 270 & 1.10 & 1 & -0.06 & 3.40 & 0.00 & 0.00\\
            5   & 10, 8 & 3, 5, 11, 2  & t & 0 & 0.13, 9.10 & 2 & 0.09 & 8.99 & 0.02 & 0.02\\
            5  & 2, 9 & 8, 3, 5, 11  & gaussian & 0 & -0.43 & 1 & -0.28 & 67.11 & 0.00 & 0.00\\
            5  & 11, 6 & 9, 8, 3, 5  & frank & 0 & 1.16 & 1 & 0.13 & 10.76 & 0.00 & 0.00\\
            5  & 5, 4 & 6, 9, 8, 3  & frank & 0 & -0.55 & 1 & -0.06 & 2.71 & 0.00 & 0.00\\
            \addlinespace
            \rowcolor{green}
            6  & 1, 3 & 5, 11, 2, 10, 7  & frank & 0 & 1.86 & 1 & 0.20 & 34.87 & 0.00 & 0.00\\
            6  & 7, 8 & 3, 5, 11, 2, 10  & joe & 90 & 1.06 & 1 & -0.03 & 1.53 & 0.00 & 0.00\\
            6  & 10, 9 & 8, 3, 5, 11, 2  & clayton & 270 & 0.06 & 1 & -0.03 & 1.34 & 0.00 & 0.00\\
            6  & 2, 6 & 9, 8, 3, 5, 11  & indep & 0 &  & 0 & 0.00 & 0.00 & 0.00 & 0.00\\
            6  & 11, 4 & 6, 9, 8, 3, 5  & indep & 0 &  & 0 & 0.00 & 0.00 & 0.00 & 0.00\\
            \addlinespace
            \rowcolor{green}
            7  & 1, 8 & 3, 5, 11, 2, 10, 7  & frank & 0 & 2.11 & 1 & 0.22 & 42.56 & 0.00 & 0.00\\
            7  & 7, 9 & 8, 3, 5, 11, 2, 10  & t & 0 & 0.66, 6.52 & 2 & 0.46 & 205.66 & 0.25 & 0.25\\
            7  & 10, 6 & 9, 8, 3, 5, 11, 2  & clayton & 180 & 0.06 & 1 & 0.03 & 1.31 & 0.00 & 0.00\\
            7  & 2, 4 & 6, 9, 8, 3, 5, 11  & indep & 0 &  & 0 & 0.00 & 0.00 & 0.00 & 0.00\\
            \addlinespace
            \rowcolor{green}
            8  & 1, 9 & 8, 3, 5, 11, 2, 10, 7  & frank & 0 & 2.02 & 1 & 0.22 & 36.25 & 0.00 & 0.00\\
            8  & 7, 6 & 9, 8, 3, 5, 11, 2, 10  & gumbel & 270 & 1.04 & 1 & -0.04 & 3.37 & 0.00 & 0.00\\
            8  & 10, 4 & 6, 9, 8, 3, 5, 11, 2  & indep & 0 &  & 0 & 0.00 & 0.00 & 0.00 & 0.00\\
            \addlinespace
            \rowcolor{green}
            9  & 1, 6 & 9, 8, 3, 5, 11, 2, 10, 7  & gumbel & 0 & 1.09 & 1 & 0.09 & 10.55 & 0.11 & 0.00\\
            9  & 7, 4 & 6, 9, 8, 3, 5, 11, 2, 10  & joe & 180 & 1.20 & 1 & 0.10 & 9.25 & 0.00 & 0.22\\
            \addlinespace
            \rowcolor{green}
            10  & 1, 4 & 6, 9, 8, 3, 5, 11, 2, 10, 7  & gumbel & 90 & 1.12 & 1 & -0.11 & 9.11 & 0.00 & 0.00\\
            \bottomrule
        \end{tabular}}
        \caption{$D$-vine tree sequence, tree $T_1$ to tree $T_{10},$ of \ref{eq: d-vine-resp-cov-est}. The variables (response and contributing factors) are labeled as in Table \ref{tab: labels}.}
        \label{tab: vine-trees}
    \end{table}

\end{appendix}

\section*{Acknowledgement} The authors would like to thank the \href{https://www.igsse.gs.tum.de/en/research/project-teams/14th-cohort/1408-copfly/}{CopFly} team at the
Institute of Flight System Dynamics of Prof. Dr.-Ing. Florian Holzapfel, Technical University of Munich, for their collaboration and insights. A special thanks goes to Xioalong Wang for his immense help for explaining the data set used in this research.

\section*{Funding} The first author acknowledges the support of \href{https://www.igsse.gs.tum.de/en/home/}{IGSSE}, and the second author is supported by the German Research Foundation.

\bibliography{references}  %%% Uncomment this line and comment out the ``thebibliography'' section below to use the external .bib file (using bibtex) .

\begin{thebibliography}{}

\bibitem[Aas, 2016]{aas2016pair}
Aas, K. (2016).
\newblock Pair-copula constructions for financial applications: A review.
\newblock {\em Econometrics}, 4(4):43.

\bibitem[Aas et~al., 2009]{PPC}
Aas, K., Czado, C., Frigessi, A., and Bakken, H. (2009).
\newblock Pair-copula constructions of multiple dependence.
\newblock {\em Insurance: Mathematics and Economics}, 44(2):182--198.

\bibitem[Ahmed et~al., 2014]{ahmed2014real}
Ahmed, M.~M., Abdel-Aty, M., Lee, J., and Yu, R. (2014).
\newblock Real-time assessment of fog-related crashes using airport weather
  data: A feasibility analysis.
\newblock {\em Accident Analysis \& Prevention}, 72:309--317.

\bibitem[Anderson, 2007]{anderson2007fundamentals}
Anderson, J. (2007).
\newblock {\em Fundamentals of Aerodynamics}.
\newblock McGraw-Hill Series in Aeronautical and. McGraw-Hill Higher Education.

\bibitem[{Arnaldo Valdés} et~al., 2018]{ARNALDOVALDES2018216}
{Arnaldo Valdés}, R.~M., {Gómez Comendador}, V.~F., {Perez Sanz}, L., and
  {Rodriguez Sanz}, A. (2018).
\newblock Prediction of aircraft safety incidents using bayesian inference and
  hierarchical structures.
\newblock {\em Safety Science}, 104:216--230.

\bibitem[Au and Beck, 2001]{AU2001263}
Au, S.-K. and Beck, J.~L. (2001).
\newblock Estimation of small failure probabilities in high dimensions by
  subset simulation.
\newblock {\em Probabilistic Engineering Mechanics}, 16(4):263--277.

\bibitem[Ayra et~al., 2019]{ayra2019bayesian}
Ayra, E.~S., R{\'\i}os~Insua, D., and Cano, J. (2019).
\newblock Bayesian network for managing runway overruns in aviation safety.
\newblock {\em Journal of Aerospace Information Systems}, 16(12):546--558.

\bibitem[Bedford and Cooke, 2001]{Cook&bedf}
Bedford, T. and Cooke, R. (2001).
\newblock Probability density decomposition for conditionally dependent random
  variables modeled by vines.
\newblock {\em Annals of Mathematics and Artificial Intigence}, 32(1):245--268.

\bibitem[Bernard and Czado, 2015]{BernCzado}
Bernard, C. and Czado, C. (2015).
\newblock Conditional quantiles and tail dependence.
\newblock {\em Journal of Multivariate Analysis}, 138:104--126.

\bibitem[Burin, 2011]{KeystoSafetyArrival}
Burin, J.~M. (2011).
\newblock Keys to a safe arrival.
\newblock \url{https://flightsafety.org/asw-article/keys-to-a-safe-arrival/},
  Last accessed on 2022-02-25.

\bibitem[ByesFusion, 2020]{GeNIe}
ByesFusion (2020).
\newblock Genie, graphical network interface.

\bibitem[Chang et~al., 2016]{chang2016human}
Chang, Y.-H., Yang, H.-H., and Hsiao, Y.-J. (2016).
\newblock Human risk factors associated with pilots in runway excursions.
\newblock {\em Accident Analysis \& Prevention}, 94:227--237.

\bibitem[Czado, 2010]{czado2010pair}
Czado, C. (2010).
\newblock Pair-copula constructions of multivariate copulas.
\newblock In {\em Copula theory and its applications}, pages 93--109. Springer.

\bibitem[Czado, 2019]{czado2019analyzing}
Czado, C. (2019).
\newblock {\em Analyzing Dependent Data with Vine Copulas: A Practical Guide
  With R}.
\newblock Lecture Notes in Statistics. Springer International Publishing.

\bibitem[Czado and Nagler, 2022]{CzadoNaglerReview}
Czado, C. and Nagler, T. (2022).
\newblock Vine copula based modeling.
\newblock {\em Annual Review of Statistics and Its Application}, 9(1):null.

\bibitem[Drees, 2016]{Drees}
Drees, L. (2016).
\newblock {\em Predictive Analysis: Quantifying Operational Airline Risks}.
\newblock Dissertation, Technische Universität München, München.

\bibitem[Drees et~al., 2014]{drees2014using}
Drees, L., Wang, C., and Holzapfel, F. (2014).
\newblock Using subset simulation to quantify stakeholder contribution to
  runway overrun.
\newblock {\em Proceedings of Probabilistic Safety Assessment and Management
  PSAM}, 12.

\bibitem[{Flight Safety Foundation}, 2009]{flight}
{Flight Safety Foundation} (2009).
\newblock Reducing the risk of runway excursions.

\bibitem[Gu and Wang, 2014]{run2014estimation}
Gu, R.-p. and Wang, P. (2014).
\newblock Estimation of wet and contaminated runway landing distance based on
  multiple linear regression.
\newblock {\em Journal of Civil Aviation University of China}, 32(3):20.

\bibitem[Hao and Naiman, 2007]{hao2007quantile}
Hao, L. and Naiman, D.~Q. (2007).
\newblock {\em Quantile regression}.
\newblock Number 149. Sage.

\bibitem[Hu et~al., 2016]{hu2016study}
Hu, C., Zhou, S.-H., Xie, Y., and Chang, W.-B. (2016).
\newblock The study on hard landing prediction model with optimized parameter
  svm method.
\newblock In {\em 2016 35th Chinese Control Conference (CCC)}, pages
  4283--4287. IEEE.

\bibitem[IATA, 2022]{IATA2024}
IATA (2022).
\newblock Air passenger numbers to recover in 2024.
\newblock \url{https://www.iata.org/en/pressroom/2022-releases/2022-03-01-01/},
  Last accessed on 2022-03-09.

\bibitem[ICAO, 2006]{ICAO2006}
ICAO (2006).
\newblock Icao annex 19, safety management.
\newblock \url{https://www.icao.int/safety/SafetyManagement/}, Last accessed on
  2022-02-25.

\bibitem[ICAO, 2013]{ICAO2013}
ICAO (2013).
\newblock {\em Safety Management Manual (SMS)}.
\newblock Number Doc 9859. Third edition edition.

\bibitem[Investigation, 2021]{Fokker100}
Investigation, A.~O. (2021).
\newblock {\em Runway overrun involving Fokker 100, VH-NHY}.

\bibitem[Jenkins and Aaron, 2012]{jenkins2012reducing}
Jenkins, M. and Aaron, R. (2012).
\newblock reducing runway landing overruns.
\newblock {\em Aero Magazine}, 3:14--19.

\bibitem[Joe, 1996]{joe1997multivariate}
Joe, H. (1996).
\newblock {\em Multivariate models and multivariate concepts}.
\newblock Chapman \& Hall.

\bibitem[Joe, 2014]{joe2014dependence}
Joe, H. (2014).
\newblock {\em Dependence modeling with copulas}.
\newblock CRC press.

\bibitem[Koenker, 2005]{koenker_2005}
Koenker, R. (2005).
\newblock {\em Quantile Regression}.
\newblock Econometric Society Monographs. Cambridge University Press.

\bibitem[Koenker, 2021]{quantregP}
Koenker, R. (2021).
\newblock {\em quantreg: Quantile Regression}.
\newblock R package version 5.86.

\bibitem[Koenker et~al., 2017]{koenker2017handbook}
Koenker, R., Chernozhukov, V., He, X., and Peng, L. (2017).
\newblock {\em Handbook of quantile regression}.
\newblock CRC press.

\bibitem[Koenker and Hallock, 2001]{koenker2001quantile}
Koenker, R. and Hallock, K.~F. (2001).
\newblock Quantile regression.
\newblock {\em Journal of economic perspectives}, 15(4):143--156.

\bibitem[Koenker and Bassett, 1978]{Koenker}
Koenker, R.~W. and Bassett, G. (1978).
\newblock Regression quantiles.
\newblock {\em Econometrica}, 46(1):33--50.

\bibitem[Kraus and Czado, 2017]{KRAUS20171}
Kraus, D. and Czado, C. (2017).
\newblock D-vine copula based quantile regression.
\newblock {\em Computational Statistics \& Data Analysis}, 110:1--18.

\bibitem[Li et~al., 2013]{li2013optimal}
Li, Q., Lin, J., and Racine, J.~S. (2013).
\newblock Optimal bandwidth selection for nonparametric conditional
  distribution and quantile functions.
\newblock {\em Journal of Business \& Economic Statistics}, 31(1):57--65.

\bibitem[McLachlan and Peel, 2000]{mclachlan2000finite}
McLachlan, G. and Peel, D. (2000).
\newblock {\em Finite Mixture Models}.
\newblock Wiley Series in Probability and Statistics. Wiley.

\bibitem[Nagler, 2021]{vinereg}
Nagler, T. (2021).
\newblock {\em vinereg: D-Vine Quantile Regression}.
\newblock R package version 0.7.4.

\bibitem[Nagler and Vatter, 2021]{rvinecopulib}
Nagler, T. and Vatter, T. (2021).
\newblock {\em rvinecopulib: High Performance Algorithms for Vine Copula
  Modeling}.
\newblock R package version 0.5.5.1.1.

\bibitem[Noh et~al., 2015]{noh2015semiparametric}
Noh, H., Ghouch, A.~E., and Van~Keilegom, I. (2015).
\newblock Semiparametric conditional quantile estimation through copula-based
  multivariate models.
\newblock {\em Journal of Business \& Economic Statistics}, 33(2):167--178.

\bibitem[Patton, 2012]{patton2012review}
Patton, A.~J. (2012).
\newblock A review of copula models for economic time series.
\newblock {\em Journal of Multivariate Analysis}, 110:4--18.

\bibitem[Rosenblatt, 1952]{Rosenblatt}
Rosenblatt, M. (1952).
\newblock Remarks on a multivariate transformation.
\newblock {\em The Annals of Mathematical Statistics}, 23(3):470--472.

\bibitem[Sch{\"o}bi and Sudret, 2019]{schobi2019global}
Sch{\"o}bi, R. and Sudret, B. (2019).
\newblock Global sensitivity analysis in the context of imprecise probabilities
  (p-boxes) using sparse polynomial chaos expansions.
\newblock {\em Reliability Engineering \& System Safety}, 187:129--141.

\bibitem[Scholz, 2012]{scholz2012aircraft}
Scholz, D. (2012).
\newblock {\em Aircraft design}.
\newblock Springer.

\bibitem[Sforza, 2014]{sforza2014commercial}
Sforza, P. (2014).
\newblock {\em Commercial Airplane Design Principles}.
\newblock Elsevier aerospace engineering series. Elsevier Science.

\bibitem[Simons, 2015]{simons2015model}
Simons, M. (2015).
\newblock {\em Model Aircraft Aerodynamics}.
\newblock Chris Lloyd Sales \& Marketing.

\bibitem[Sklar, 1959]{Skla59}
Sklar, A. (1959).
\newblock Fonctions de r\'epartition \`a n dimensions et leurs marges.
\newblock {\em Publications de l'Institut de Statistique de l'Universit\'e de
  Paris}, 8:229--231.

\bibitem[Spokoiny et~al., 2013]{spokoiny2013local}
Spokoiny, V., Wang, W., and H{\"a}rdle, W.~K. (2013).
\newblock Local quantile regression.
\newblock {\em Journal of Statistical Planning and Inference},
  143(7):1109--1129.

\bibitem[Tang et~al., 2015]{tang2015copula}
Tang, X.-S., Li, D.-Q., Zhou, C.-B., and Phoon, K.-K. (2015).
\newblock Copula-based approaches for evaluating slope reliability under
  incomplete probability information.
\newblock {\em Structural Safety}, 52:90--99.

\bibitem[Vald{\'e}s et~al., 2011]{valdes2011development}
Vald{\'e}s, R. M.~A., Comendador, F.~G., Gord{\'u}n, L.~M., and Nieto, F. J.~S.
  (2011).
\newblock The development of probabilistic models to estimate accident risk
  (due to runway overrun and landing undershoot) applicable to the design and
  construction of runway safety areas.
\newblock {\em Safety science}, 49(5):633--650.

\bibitem[Wagner and Barker, 2014]{wagner2014statistical}
Wagner, D.~C. and Barker, K. (2014).
\newblock Statistical methods for modeling the risk of runway excursions.
\newblock {\em Journal of Risk Research}, 17(7):885--901.

\bibitem[Wang et~al., 2014a]{wang2014quantification}
Wang, C., Drees, L., Gissibl, N., H{\"o}hndorf, L., Sembiring, J., and
  Holzapfel, F. (2014a).
\newblock Quantification of incident probabilities using physical and
  statistical approaches.
\newblock In {\em 6th International Conference on Research in Air
  Transportation. Istanbul, Turkey}.

\bibitem[Wang et~al., 2014b]{wang2014analysis}
Wang, L., Wu, C., and Sun, R. (2014b).
\newblock An analysis of flight quick access recorder (qar) data and its
  applications in preventing landing incidents.
\newblock {\em Reliability Engineering \& System Safety}, 127:86--96.

\bibitem[Wang et~al., 2020]{PCE}
Wang, X., Fang, X., Beller, L., and Holzapfel, F. (2020).
\newblock Calibration of contributing factors for model-based predictive
  analysis algorithm using polynomial chaos expansion methods.

\bibitem[Wong et~al., 2006]{wong2006quantifying}
Wong, D.~K., Pitfield, D., Caves, R.~E., and Appleyard, A. (2006).
\newblock Quantifying and characterising aviation accident risk factors.
\newblock {\em Journal of Air Transport Management}, 12(6):352--357.

\bibitem[You et~al., 2013]{you2013effects}
You, X., Ji, M., and Han, H. (2013).
\newblock The effects of risk perception and flight experience on airline
  pilots’ locus of control with regard to safety operation behaviors.
\newblock {\em Accident Analysis \& Prevention}, 57:131--139.

\bibitem[Zhao and Zhang, 2022]{zhao2022research}
Zhao, N. and Zhang, J. (2022).
\newblock Research on the prediction of aircraft landing distance.
\newblock {\em Mathematical Problems in Engineering}, 2022.

\bibitem[Zwirglmaier and Straub, 2016]{zwirglmaier2016discretization}
Zwirglmaier, K. and Straub, D. (2016).
\newblock A discretization procedure for rare events in bayesian networks.
\newblock {\em Reliability Engineering \& System Safety}, 153:96--109.

\end{thebibliography}

\end{document}